\documentclass[twocolumn, floatfix, tighten, fleqn, useAMS, usenatbib]{aastex63}
\acceptjournal{Universe}

\RequirePackage{silence}
\usepackage{float}
\usepackage{amsmath}	% Advanced maths commands
\usepackage{amssymb}	% Extra maths symbols

\usepackage{amsfonts}
\usepackage{graphicx}	% Including figure files
\usepackage{bm}		% Bold maths symbols, including upright Greek
\usepackage{mathtools}
\usepackage{booktabs}
\usepackage{listings}
\usepackage{color}
\usepackage{threeparttable}
\usepackage{multirow}
\usepackage[perpage,symbol*,flushmargin]{footmisc}%At end of square bracket: ,symbol* \usepackage[T1]{fontenc}
\usepackage{ae,aecompl}
\usepackage{array}
\usepackage{soul}

\usepackage[T1]{fontenc}
\usepackage[perpage,symbol*]{footmisc}
\usepackage{ae,aecompl}

\usepackage{substitutefont}
\substitutefont{TS1}{aer}{cmr}

\let\oldtextbf=\textbf
\renewcommand\textbf[1]{{\boldmath\oldtextbf{#1}}}

\begin{document}
%\label{firstpage} %AAS does not use this, or the corresponding lastpage command.
\title{The Magellanic Clouds are very rare in the IllustrisTNG simulations}
\shorttitle{The Magellanic Clouds are very rare in $\Lambda$CDM} %\Compulsory, 40-45 characters, editors will use discretion.
%\maketitle command is illegal.

\shortauthors{M. Haslbauer, I. Banik, P. Kroupa, H. Zhao \& E. Asencio} %Compulsory.

\author[0000-0002-5101-6366]{Moritz Haslbauer}
\affiliation{Helmholtz-Institut f\"ur Strahlen- und Kernphysik (HISKP), University of Bonn, Nussallee 14$-$16, D-53115 Bonn, Germany}
%\email{mhaslbauer@astro.uni-bonn.de}

\author[0000-0002-4123-7325]{Indranil Banik*}
\affiliation{Scottish Universities Physics Alliance, University of Saint Andrews, North Haugh, Saint Andrews, Fife, KY16 9SS, UK}
\affiliation{Institute of Cosmology \& Gravitation, University of Portsmouth, Dennis Sciama Building, Burnaby Road, Portsmouth PO1 3FX, UK}
\email{ib45@st-andrews.ac.uk}

\author[0000-0002-7301-3377]{Pavel Kroupa}
\affiliation{Helmholtz-Institut f\"ur Strahlen- und Kernphysik (HISKP), University of Bonn, Nussallee 14$-$16, D-53115 Bonn, Germany}
\affiliation{Astronomical Institute, Faculty of Mathematics and Physics, Charles University, V Hole\v{s}ovi\v{c}k\'ach 2, CZ-180 00 Praha 8, Czech Republic}

\author[0000-0001-9715-8688]{Hongsheng Zhao}
\affiliation{Scottish Universities Physics Alliance, University of Saint Andrews, North Haugh, Saint Andrews, Fife, KY16 9SS, UK} 

\author[0000-0002-3951-8718]{Elena Asencio}
\affiliation{Helmholtz-Institut f\"ur Strahlen- und Kernphysik (HISKP), University of Bonn, Nussallee 14$-$16, D-53115 Bonn, Germany}

\begin{abstract} %One paragraph only, no blank line below.
The Large and Small Magellanic Cloud (LMC and SMC) form the closest interacting galactic system to the Milky Way, therewith providing a laboratory to test cosmological models in the local Universe. We quantify the likelihood for the Magellanic Clouds (MCs) to be observed within the $\Lambda$CDM model using hydrodynamical simulations of the IllustrisTNG project. The orbits of the MCs are constrained by proper motion measurements taken by the \emph{Hubble Space Telescope} and \emph{Gaia}. The MCs have a mutual separation of $d_{\mathrm{MCs}}~=~24.5\,\mathrm{kpc}$ and a relative velocity of $v_{\mathrm{MCs}}~=~90.8\,\mathrm{km\,s^{-1}}$, implying a phase-space density of $f_{\mathrm{MCs,obs}}~\equiv~(d_{\mathrm{MCs}} \cdot v_{\mathrm{MCs}})^{-3}~=~9.10\times10^{-11}\,\mathrm{km^{-3}\,s^{3}\,kpc^{-3}}$. We select analogues to the MCs based on their stellar masses and distances in MW-like halos. None of the selected LMC analogues have a higher total mass and lower Galactocentric distance than the LMC, resulting in $>3.75\sigma$ tension. We also find that the $f_{\mathrm{MCs}}$ distribution in the highest resolution TNG50 simulation is in $3.95\sigma$ tension with observations. Thus, a hierarchical clustering of two massive satellites like the MCs in a narrow phase-space volume is unlikely in $\Lambda$CDM, presumably because of short merger timescales due to dynamical friction between the overlapping dark matter halos. We show that group infall led by an LMC analogue cannot populate the Galactic disc of satellites (DoS), implying that the DoS and the MCs formed in physically unrelated ways in $\Lambda$CDM. Since the $20^\circ$ alignment of the LMC and DoS orbital poles has a likelihood of $P=0.030$ ($2.17\sigma$), adding this $\chi^2$ to that of $f_{\mathrm{MCs}}$ gives a combined likelihood of $P = 3.90\times10^{-5}$ ($4.11\sigma$).
\end{abstract}

%see http://astrothesaurus.org/thesaurus/alphabetical-browse/
%Unified Astronomy Thesaurus concepts:
\keywords{Magellanic Clouds (990); Milky Way Galaxy (1054); Galaxy interactions (600); Galaxy mergers (608); Dynamical friction (422); Cold dark matter (265)}

\section{Introduction}
\label{sec:Introduction}

The Large Magellanic Cloud (LMC) and Small Magellanic Cloud (SMC) are the most massive and some of the closest satellite galaxies of the Milky Way (MW). The gravitational interaction between these three galaxies makes the nearby Universe an interesting laboratory to test galaxy formation and evolution models and gravitational theories. In particular, the proximity of the MW, LMC, and SMC means that their hypothetical dark matter halos are strongly overlapping such that Chandrasekhar dynamical friction ought to be pronounced \citep{Kroupa_2015}. This has been argued to prevent their observed configuration from arising in the dark matter framework given the essential constraint that the Magellanic Clouds (MCs) must have experienced at least one close encounter with each other $1-4$~Gyr ago to develop the Magellanic Stream \citep[MS;][]{Oehm_2024}, while the synchronised star formation histories of both galaxies \citep{Massana_2022} suggests that the MCs had four close encounters over the past 3~Gyr. Such tests require precise proper motion measurements in order to constrain the orbits of the MCs. Historically, the first proper motion measurements of the MCs were reported by \citet{Kroupa_1994} using the Positions and Proper Motions \citep[PPM;][]{Roesser_1993} catalogue. The \emph{High Precision Parallax Collecting Satellite} \citep[\emph{Hipparcos};][]{Perryman_1997} was the first high-precision astrometry space mission, allowing further improvement in the proper motions of the MCs \citep{Kroupa_1997}. The precision of the proper motion measurements improved further thanks to the \emph{Hubble Space Telescope} \citep[\emph{HST}; e.g.][]{Kallivayalil_2006a, Kallivayalil_2006b, Kallivayalil_2013} and recently \emph{Gaia} \citep{GaiaCollaboration_2016, Gaia_2018a, Gaia_2018b}. 

In this contribution, we aim to quantify the likelihood of MW-LMC-SMC triple systems occurring in the Lambda-Cold Dark Matter ($\Lambda$CDM) cosmological model \citep{Efstathiou_1990, Ostriker_1995}. Our analysis relies on the stellar masses and orbits of the MCs using proper motion measurements taken by the \emph{HST} \citep{Kallivayalil_2013} and \emph{Gaia} Data Release 2 \citep[\emph{Gaia}~DR2;][]{GaiaCollaboration_2016, Gaia_2018a, Gaia_2018b}. The LMC and SMC have a stellar mass of $M_{\star} = 1.5 \times 10^9\,M_\odot$ and $M_{\star} = 4.6 \times 10^8\,M_\odot$, respectively \citep{McConnachie_2012}. We obtain the 3D position and velocity vectors of the MCs in Galactic coordinates from \citet{Pawlowski_2020}, who combined proper motion data from the \emph{HST} \citep{Kallivayalil_2013} with \emph{Gaia}~DR2 \citep{Gaia_2018a, Gaia_2018b}. The LMC and SMC are located at Galactocentric distances of $50.0 \,\mathrm{kpc}$ and $61.3 \,\mathrm{kpc}$, respectively, with a mutual separation of only $24.5\,\mathrm{kpc}$. The relative velocity of the MCs is $90.8\,\mathrm{km\,s^{-1}}$, which is quite small given that the LMC and SMC have Galactocentric velocities of $323.8 \,\mathrm{km\,s^{-1}}$ and $245.6 \,\mathrm{km\,s^{-1}}$, respectively, with the LMC ahead of the SMC in their orbit around the MW. Assuming as the null hypothesis that the $\Lambda$CDM paradigm is the correct description of the Universe, the rather rapid motion of the LMC combined with other observations of the MCs (discussed further below) implies that the MCs are most likely on their first pericentre passage past the MW \citep{Besla_2007}. As a result, the fact that the LMC is leading the SMC is not a strong constraint $-$ slight differences in the infall time and pre-infall trajectory can mask the $\approx 10\times$ larger expected dynamical friction on the LMC, which moreover could actually speed up if its orbit decayed more due to the so-called `donkey effect'. The above-described geometrical and kinematic configuration of the MW-MCs system causes some interesting effects, which can be used as a test of cosmological and gravitational theories.

Besides the kinematic data, the evolutionary history of the MCs can also be accessed by studying their environment and internal properties over cosmic time. Strong evidence that the LMC and SMC interacted over at least the last 4~Gyr is provided by their star formation histories (SFHs). In particular, the SFH of the SMC has two peaks at lookback times of $\approx 9$~Gyr and $\approx 4.5$~Gyr \citep{Weisz_2013}, while the highest star formation rate (SFR) of the LMC occurred $0.5-4$~Gyr ago \citep{Mazzi_2021}. The SFRs of both MCs increased significantly over the last 3.5~Gyr \citep[see e.g. figure~5 of][]{Weisz_2013}. This and the correlation between the SFHs of the LMC and SMC over the last 3.5~Gyr \citep[see figure~2 of][]{Massana_2022} are all suggestive of star formation triggered by a mutual interaction.

The MCs are surrounded by the MS, a gaseous structure with leading and trailing arms that extend over about $200^{\circ}$ on our sky \citep{Nidever_2010}. The MS also aligns with the Disk of Satellites (DoS): the MS normal and the DoS normal have an angular distance of $24^{\circ}$ \citep{Pawlowski_2012a}. Different MS formation processes have been proposed \citep[for a review, see][]{Donghia_2016}. Currently, the most preferred scenario is that the MS is a tidal tail formed mostly through gravitational interactions between the MCs, though with ram-pressure stripping from hot gas surrounding the MW and tides raised by its disc also playing a role in the context of a first infall scenario \citep{Hammer_2015}. However, this would not explain the alignment of the MCs and the MS with the satellite plane around the MW (the DoS; see Section~\ref{subsec:MCs and VPoS}).

The formation and evolution of the MCs in cosmological $\Lambda$CDM simulations was investigated in previous studies \citep[e.g.,][]{Busha_2011, BoylanKolchin_2011, SantosSantos_2021} $-$ we discuss this further in Section~\ref{sec:Discussion}. We readdress the MCs in the $\Lambda$CDM context by recalculating the likelihood of MW-like galaxies hosting analogues to the MCs using the state-of-the-art hydrodynamical simulation runs of the Illustris The Next Generation (hereafter TNG) project \citep{Pillepich_2018, Nelson_2019}. In addition to this major advance on the theoretical side, we also take into account the geometrical configuration and kinematics of the MCs using proper motion measurements obtained with the \emph{HST} \citep{Kallivayalil_2013} and from \emph{Gaia}~DR2 \citep{GaiaCollaboration_2016, Gaia_2018a, Gaia_2018b}. 
%Here, we calculate the likelihood of MW-like galaxies hosting analogues to the MCs using the hydrodynamical simulation runs of the Illustris The Next Generation (hereafter TNG) project \citep[][]{Pillepich_2018, Nelson_2019}.

The layout of this paper is as follows: Section~\ref{sec:Methods} describes the TNG simulation runs and the selection criteria used to define analogues to the MW and the MCs. The likelihood of the MCs within the $\Lambda$CDM framework is quantified in Section~\ref{sec:Results}. This section also includes an investigation of the formation of the MS in a cosmological context by tracing individual analogues to the MCs back through cosmic time. In Section~\ref{sec:Discussion}, we compare our results with previous studies on the occurrence rate of the MCs in the $\Lambda$CDM framework and discuss the role of dynamical friction on the orbits of the MCs. We also consider the MCs in the broader context of the Local Group (LG). Our concluding remarks are given in Section~\ref{sec:Conclusion}.

\section{Methods}
\label{sec:Methods}

In this section, we first summarize the observational properties of the MCs. We then introduce the hydrodynamical cosmological TNG simulations and state our adopted selection criteria for analogues to the MW and the MCs.

\subsection{Observations}
\label{subsec:Observations}

\begin{table*}
  \centering
	\begin{tabular}{lllllllllll}
	\hline
    Object & $M_{\star}$ & $r_x$ & $r_y$ & $r_z$ & $d$ & $v_x$ & $v_y$ & $v_z$ & $v_{\mathrm{tot}}$& $h/10^3$ \\  
    & [$M_\odot$] & [kpc] & [kpc] & [kpc]  & [kpc] & [$\mathrm{km\,s^{-1}}$] & [$\mathrm{km\,s^{-1}}$] & [$\mathrm{km\,s^{-1}}$] & [$\mathrm{km\,s^{-1}}$] & [$\mathrm{kpc\,km\,s^{-1}}$] \\ \hline  
    LMC & $1.5 \times 10^9$ & $-0.6$ & $-41.8$ & $-27.5$ & $50.0$ & $-42 \pm 6$ & $-223 \pm 4$ & $231 \pm 4$ & $323.8$ & $15.9$ \\
    SMC & $4.6 \times 10^8$ & $~16.5$ & $-38.5$ & $-44.7$ & $61.3$ & $\quad~ 6 \pm 8$ & $-180 \pm 7$ & $167 \pm 6$ & $245.6$ & $15.0$ \\ 
	\hline
	\end{tabular}
	\caption{Observational properties of the MCs based on tables~1 and 2 of \citet{Pawlowski_2020}. The proper motions are a combination of data from \emph{Gaia}~DR2 \citep{Gaia_2018b} and the \emph{HST} \citep{Kallivayalil_2013}, as discussed further in section~2.1 of \citet{Pawlowski_2020}. From left to right: stellar mass; position $\bm{r}$ in Galactocentric Cartesian coordinates; distance to the Galactic centre; velocity $\bm{v}$ in Galactocentric Cartesian coordinates; total velocity wrt. the MW; and absolute specific angular momentum $h \equiv \lvert \bm{r} \times \bm{v} \rvert$. The stellar masses of the LMC and SMC are taken from table~4 of \citet{McConnachie_2012}.}
  \label{tab:observations}
\end{table*}

The 3D positions and velocities of the MCs in Galactocentric Cartesian coordinates are summarized in Table~\ref{tab:observations} and are taken from table~2 of \citet{Pawlowski_2020}, whose results are derived by combining proper motion measurements from \emph{Gaia}~DR2 \citep{Gaia_2018b} with those from the \emph{HST} \citep{Kallivayalil_2013}. For a detailed description of these measurements, we refer the reader to section~2.1 of \citet{Pawlowski_2020}.

The phase-space configuration of the MW-MCs system is studied by combining the phase-space density of the MCs with a measure of the LMC being close and moving fast wrt. the MW. We focus mostly on the specific phase-space density of the MCs, which is given by
\begin{eqnarray}
    f_{\mathrm{MCs}} ~&\equiv&~ \left( d_{\mathrm{LMC-SMC}} \cdot v_{\mathrm{LMC-SMC}} \right)^{-3} \, , ~~ \text{with} \label{eq:specific_phase_space_density} \\     
    f_{\mathrm{MCs, obs}} ~&\approx&~ 9.10~\times~10^{-11}~\,~\mathrm{km^{-3}\,s^{3}\,kpc^{-3}}
    \label{eq:specific_phase_space_density_obs}
\end{eqnarray}
being the observationally inferred value. The distances to the MCs are very well known by now, so the uncertainty in $f_{\mathrm{MCs, obs}}$ comes almost entirely from the $\approx 15 \, \mathrm{km\,s^{-1}}$ uncertainty in their relative velocity. Since this is a fractional uncertainty of $\approx 1/6$, it is clear that $f_{\mathrm{MCs, obs}}$ has an uncertainty of about 50\% or 0.2~dex. However, since proper motion uncertainties are more likely to inflate the measured relative velocity and thereby reduce $f_{\mathrm{MCs, obs}}$, we take the conservative approach of trying to match its observed value without taking into account its uncertainty.

The geometrical and kinematic configuration of the MW-LMC system is quantified by introducing the inverse kinematic timescale
\begin{eqnarray}
    \Omega_{\mathrm{LMC}} ~\equiv~ \frac{v_{\mathrm{MW-LMC}}}{d_{\mathrm{MW-LMC}}} \, ,
    \label{eq:velocity_distance_LMC}
\end{eqnarray}
which is also a measure related to the angular velocity. Observationally, $\Omega_{\mathrm{LMC,obs}} = 6.48 \, \mathrm{km\,s^{-1}\,kpc^{-1}}$ or (151~Myr)\textsuperscript{$-1$}. The parameters $f_{\mathrm{MCs,obs}}$ and $\Omega_{\mathrm{LMC,obs}}$ contain unrelated information as the former relates to the LMC-SMC system without regard to the MW, while the latter refers to the MW-LMC system without regard to the SMC. $f_{\mathrm{MCs}}$ and $\Omega_{\mathrm{LMC}}$ are thus independent. The parameters also do not include the masses of the MCs or the MW because these are already considered by the stellar mass selection criteria applied to our initial samples in Sections~\ref{subsec:Selection criteria} and \ref{subsec:Selection criteria of MCs-analogues}. In Section~\ref{subsubsec:Total mass of LMC analogues}, we will consider if the masses of the LMC analogues are compatible with other constraints on the $\Lambda$CDM framework.

\subsection{The TNG cosmological simulations}
\label{subsec:Cosmological simulation}

The IllustrisTNG project \citep{Pillepich_2018, Pillepich_2019, Nelson_2018, Nelson_2019, Nelson_2019_TNG50, Marinacci_2018, Naiman_2018, Springel_2018} is a further development of the original Illustris project \citep{Nelson_2014, Nelson_2015}, a set of hydrodynamical cosmological simulations performed within the $\Lambda$CDM framework. The different simulation runs of the TNG project are consistent with the Planck-2015 results as they use a global Hubble constant of $H_{0} = 67.74\,\mathrm{km\,s^{-1}\,Mpc^{-1}}$ and a present-day baryonic matter, total matter, and dark energy density in units of the cosmic critical density of $\Omega_{\mathrm{b},0} = 0.0486$, $\Omega_{\mathrm{m},0} = 0.3089$, and $\Omega_{\Lambda,0} = 0.6911$, respectively \citep{Planck_2016_IllustrisTNG}. 

In order to quantify the likelihood of LMC and SMC analogues in the $\Lambda$CDM framework, we employ the TNG50-1, TNG100-1, and TNG300-1 simulation runs. These have cubic box sides of $35 \, h^{-1} = 51.7$ (TNG50-1), $75 \, h^{-1} = 110.7$ (TNG100-1), and $205 \, h^{-1} = 302.6$ (TNG300-1) comoving Mpc (cMpc), respectively, where $h$ is the present Hubble constant $H_0$ in units of $100\,\mathrm{km\,s^{-1}\,Mpc^{-1}}$. TNG300-1 has the largest simulation volume within the IllustrisTNG project, which is important to build up sufficient statistics. The TNG50-1 run has the smallest simulation volume but is the highest mass and spatial resolution realization, with a baryonic element mass of $m_{\mathrm{b}} = 8.5 \times 10^{4} \,M_\odot$ and a dark matter particle mass of $m_{\mathrm{dm}} = 4.5 \times 10^{5}\,M_\odot$. TNG100-1 (TNG300-1) has $m_{\mathrm{b}} = 1.4 \times 10^{6} \,M_\odot$ ($m_{\mathrm{b}} = 1.1 \times 10^{7} \,M_\odot$) and $m_{\mathrm{dm}} = 7.5 \times 10^{6}\,M_\odot$ ($m_{\mathrm{dm}} = 5.9 \times 10^{7}\,M_\odot$). We also use the lower resolution realizations of the TNG50 and TNG100 runs, i.e. TNG50-2 ($m_{\mathrm{b}} = 6.8 \times 10^{5} \,M_\odot$, $m_{\mathrm{dm}} = 3.6 \times 10^{6} \,M_\odot$), TNG50-3 ($m_{\mathrm{b}} = 5.4 \times 10^{6} \,M_\odot$, $m_{\mathrm{dm}} = 2.9 \times 10^{7} \,M_\odot$), and TNG100-2 ($m_{\mathrm{b}} = 1.1 \times 10^{7} \,M_\odot$, $m_{\mathrm{dm}} = 6.0 \times 10^{7} \,M_\odot$); to study the numerical convergence of the results presented in Section~\ref{subsubsec:Phase-space density}.

The TNG simulations self-consistently evolve the baryonic\footnote{stellar particles and gas cells} components and dark matter particles from redshift $z=127$ down to $z = 0$ (the present time), yielding 100 snapshots in the redshift range $0 \le z \le 20.05$. Halos and subhalos (substructures within halos) were identified with a standard friends-of-friends (FoF) algorithm and the \textsc{subfind} algorithm, respectively \citep{Springel_2001, Dolag_2009}. Their physical properties are listed in the FoF halo and subhalo catalogues, respectively. These can be downloaded from the IllustrisTNG website.\footnote{\url{https://www.tng-project.org/data/docs/specifications/} [15.03.2021]} We trace galaxies through cosmic time using the Illustris \textsc{sublink} merger tree catalogue \citep{RodriguezGomez_2015}. Throughout the analysis, we calculate the velocity difference between two galaxies based on their total velocities rather than the peculiar velocities listed in the online catalogues. Therefore, according to the Hubble law, the relative velocity $\bm{v}_{\mathrm{12}}$ between two objects with relative peculiar velocity $\bm{v}_{\mathrm{pec, 12}}$ is
\begin{eqnarray}
    \bm{v}_{\mathrm{12}} ~\equiv~ \bm{v}_{\mathrm{pec, 12}} + H \left( z \right) \bm{d}_{\mathrm{12}}\, ,
    \label{eq:velocities}
\end{eqnarray}
where $H \left( z \right)$ is the Hubble parameter at redshift $z$, at which time $\bm{d}_{\mathrm{12}}$ is the physical separation between the objects under consideration.

\subsection{Selecting analogues to the MW}
\label{subsec:Selection criteria}

In order to select analogues to the MW and the MCs, we use a modified version of the publicly available searching algorithm developed by \citet{Banik_2021}, who used a one-level tree code to speed up the calculations.\footnote{\url{https://seafile.unistra.fr/d/6b09464443da478d8926/} [07.11.2022]}

The virial mass of the MW is $M_{200} = \left( 1.3 \pm 0.3 \right) \times 10^{12} \, M_{\odot}$ \citep{McMillan_2017, Posti_2019}, with more recent measurements favouring values near the lower end of this range \citep{Vasiliev_2021, Kravtsov_2024}. We allow a wide range of $M_{200}$ to improve the statistics. Thus, we require that an MW-like galaxy has to be within an FoF group whose total virial mass lies in the range $0.5 \times 10^{12} < M_{200}/M_{\odot} < 2.5 \times 10^{12}$, where $M_{200}$ is the total mass of the group within a sphere whose mean density is $200\times$ the cosmic critical density.\footnote{The corresponding range of virial radii is $168-287$~kpc.} The upper limit to $M_{200}$ is designed to exclude galaxy groups and clusters.
%The MW has a virial mass of $M_{200} = 2.43 \times 10^{12} \, M_{\odot}$, with a minimum value of $0.8 \times 10^{12} \, M_{\odot}$ at the 95\% confidence level \citep{Li_2008}.

From the so-obtained FoF halo sample, we select only those which have a central subhalo with stellar mass $M_{\star} > 5 \times 10^{10}\,M_{\odot}$. This limit is roughly the $1\sigma$ lower limit of the observed stellar mass of the MW \citep[$M_{\star} = \left(6.08 \pm 1.14\right) \times 10^{10} \, M_{\odot}$;][]{Licquia_2015}.

In addition to its internal properties, an MW-like galaxy has to fulfil certain isolation criteria. Firstly, we select only MW-like galaxies which do not have a more massive halo (in terms of the virial mass $M_{200}$) within 0.5~Mpc. This is approximately the distance to M31 \citep{McConnachie_2012}, which is slightly more massive than the MW. This criterion removes especially massive interacting galaxies. Secondly, we require that there should be no further galaxy with $5\times$ the $M_{200}$ of the selected MW-like galaxy within $3 \, \mathrm{Mpc}$. This is because the nearest observed galaxy which is more massive than the MW or M31 is NGC 5128 (Centaurus A), located at a distance of $\approx 3.7 \,\mathrm{Mpc}$ \citep[see, e.g., table~1 of][]{Karachentsev_2018}. The M81 and IC~342 groups are also over 3~Mpc from the MW. For further discussion of the appropriate isolation criteria, we refer the reader to \citet{Banik_2021}.

These selection criteria give $86$ (TNG50-1), $398$ (TNG100-1), and $2117$ (TNG300-1) MW-like galaxies at $z = 0$. To increase our final sample size, we also search for analogues to the MW and the MCs (Section~\ref{subsec:Selection criteria of MCs-analogues}) up to redshift $z = 0.26$, which corresponds to a lookback time of about $3.1\,\mathrm{Gyr}$. This allows us to consider 20 snapshots, which together yield 1547, 7360, and 40075 MW-like galaxies with $0.0 \le z \le 0.26$ in the TNG50-1, TNG100-1, and TNG300-1 runs, respectively. The statistics of analogues to the MCs should be similar over this period as it is a small fraction of the Hubble time. The minimum timestep between the here considered snapshots is $120\,\mathrm{Myr}$, which is quite long given that we are trying to find snapshots where the MCs are at a rather short-lived phase of their orbit (close to pericentre). Thus, although a selected subhalo could in principle be a progenitor/descendent of a selected subhalo in a previous/subsequent snapshot, we treat all selected subhalos independently of each other. We will see later that we do have one case of the same system satisfying the selection criteria at different times, but argue that this is a reasonable way to build up the statistics because it captures how the system in question would resemble the observed configuration for a greater period of time.

\subsection{Selecting analogues to the MCs}
\label{subsec:Selection criteria of MCs-analogues}

In order to identify analogues to the MCs, we first extract and rank in $M_{\star}$ all non-central subhalos with a non-zero SubhaloFlag parameter within 250~kpc of an MW-like galaxy (Section~\ref{subsec:Selection criteria}).\footnote{A description of the SubhaloFlag parameter can be found here: \url{https://www.tng-project.org/data/docs/background/\#subhaloflag} [08.03.2022]} The restriction on the SubhaloFlag parameter excludes subhalos with a non-cosmological origin, e.g. baryonic fragments of the disc or galactic substructures.\footnote{Many additional non-cosmological subhalos are in any case removed by the $M_{\mathrm{total}}/M_{\star}$ cut as it excludes dark matter-deficient galaxies like tidal dwarfs \citep{Ploeckinger_2018, Haslbauer_2019}.}

Secondly, after ranking the subhalos according to their stellar mass, an analogue to the LMC (SMC) is defined as the most (second-most) massive satellite (in terms of $M_{\star}$) if it has a total-to-stellar mass ratio of $M_{\mathrm{total}}/M_{\star} > 5$ and $M_{\star} \ge 1.5 \times 10^{9} \, M_\odot$ ($M_{\star} \ge 4.6 \times 10^{8} \, M_\odot$) \citep[see table~4 of][]{McConnachie_2012}. The here-used lower limit on the LMC's stellar mass is $1.8\times$ smaller than the $2.7 \times 10^{9}\,M_\odot$ reported by \citet{Livio_2011}, which is often quoted in the literature. The lower value is applied in order to increase the number of LMC analogues and because the study by \citet{McConnachie_2012} appeared more recently. Note that $M_{\mathrm{total}}$ is defined as the total mass of all member baryonic and dark matter particles/gas cells which are bound to the subhalo. The minimum $M_{\mathrm{total}}/M_{\star}$ cut removes especially substructures within the MW-like galaxy.

These selection criteria give $331$ (TNG50-1), $1414$ (TNG100-1), and $3823$ (TNG300-1) LMC analogues and $343$ (TNG50-1), $766$ (TNG100-1), and $1148$ (TNG300-1) SMC analogues with $z \le 0.26$. Requiring that an MW-like galaxy has to host an analogue to both the LMC and the SMC gives 147 (TNG50-1), 454 (TNG100-1), and 601 (TNG300-1) analogues. We refer to these as our initial MCs samples because only selection criteria on the total and stellar mass components are applied. A detailed comparison with the observed MCs including different physical properties is presented in the following sections.

\section{Results}
\label{sec:Results}

In this section, we quantify the likelihood of analogues to the MCs in the $\Lambda$CDM framework by applying different observational constraints to the initial samples that we defined in Section~\ref{subsec:Selection criteria of MCs-analogues}. Individual analogues to the MCs are traced back through cosmic time in order to investigate their evolution and the formation of the MS in a cosmological context.

The frequency $P$ of analogues to the MCs around MW-like galaxies is given by the number of selected analogues divided by the number of MW-like galaxies (Section~\ref{subsec:Selection criteria}). This $p$-value is converted into an equivalent number of standard deviations for a single Gaussian variable $\chi$ by solving
\begin{eqnarray}
    1 - \frac{1}{\sqrt{2 \mathrm{\pi}}} \int_{-\chi}^\chi \exp \left( -\frac{x^2}{2} \right) \, dx ~\equiv~ P \, .
    \label{P_chi}
\end{eqnarray}
We solve this iteratively using the Newton-Raphson algorithm.

\subsection{Phase-space density of the MCs} \label{subsubsec:Phase-space density}

We select only systems with a geometrical configuration similar to the observed MW-MCs system by requiring $50 \, \mathrm{kpc} \le d_{\mathrm{MW-LMC}} \le 100\,\mathrm{kpc}$. The lower limit is applied because the LMC is currently at pericentre with a Galactocentric distance of 50~kpc. This typically removes MW-LMC systems with rather high values of $\Omega_{\mathrm{LMC}}$, which can more easily exceed the observed $\Omega_{\mathrm{LMC,obs}} = 6.48 \, \mathrm{km\,s^{-1}\,kpc^{-1}}$ for LMC analogues at smaller distances. The upper distance limit is set in order to select only LMC analogues reasonably close to their MW-like host galaxies. These additional criteria result in sample sizes of 46 (TNG50-1), 118 (TNG100-1), and 191 (TNG300-1), which in total gives 355 MCs analogues across all three TNG runs. 

\begin{figure*}
    \includegraphics[width=5.8cm]{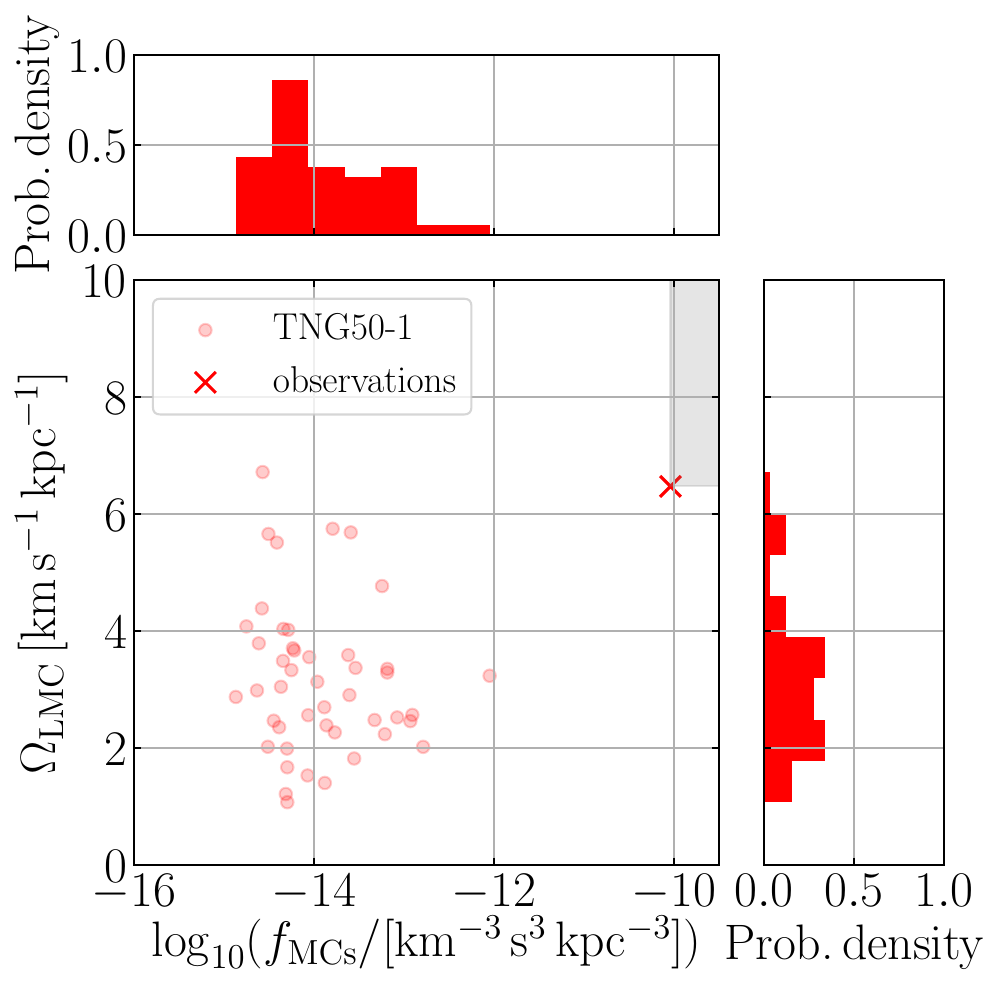}
    \includegraphics[width=5.8cm]{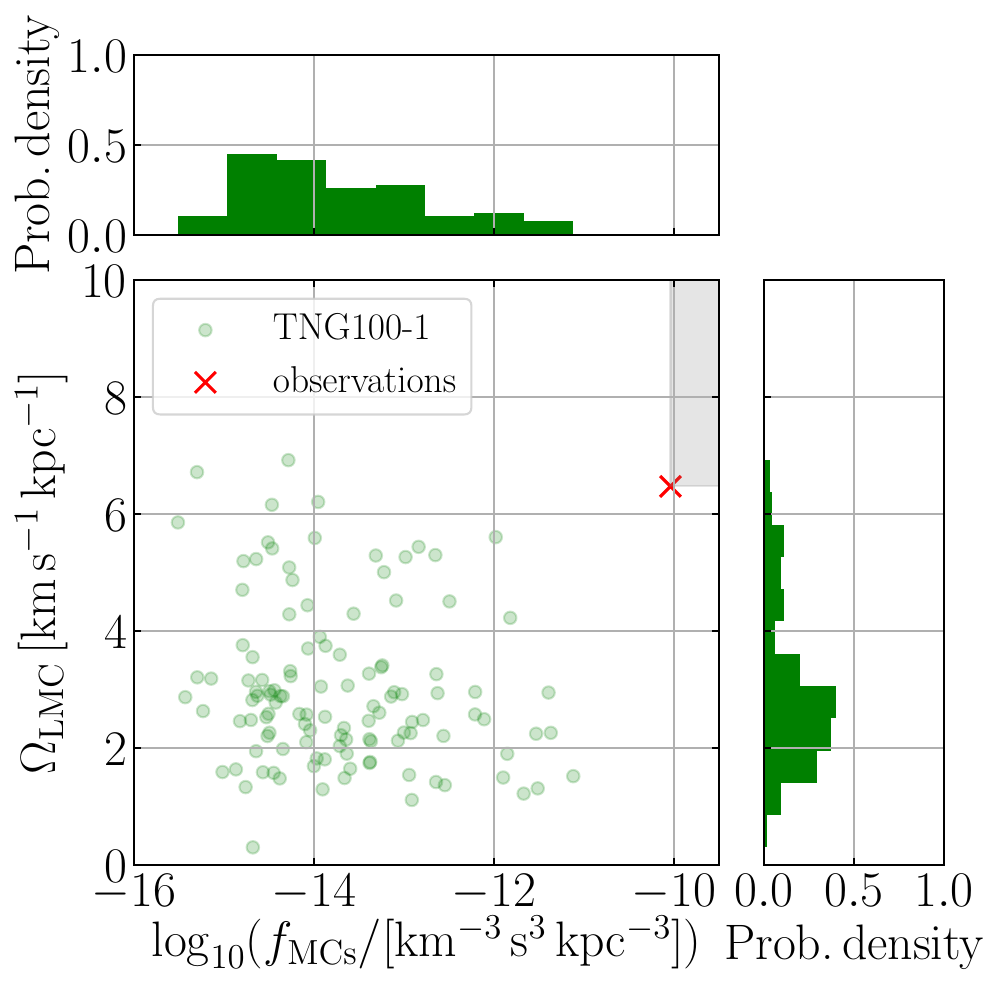}
    \includegraphics[width=5.8cm]{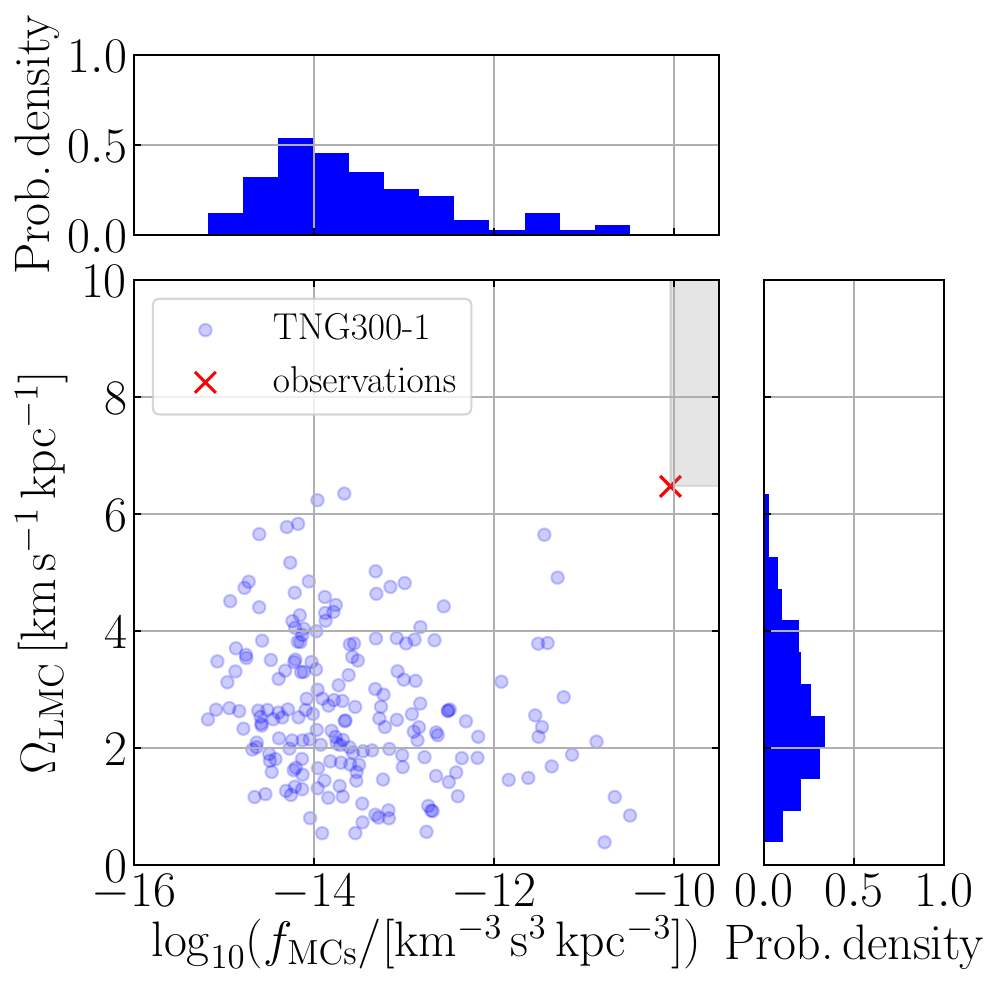}
    \caption{Distribution of the specific phase-space density of the MCs ($f_{\mathrm{MCs}}$; Equation~\ref{eq:specific_phase_space_density}) and the inverse kinematic timescale of the LMC ($\Omega_{\mathrm{LMC}}$; Equation~\ref{eq:velocity_distance_LMC}) for systems with $50 \, \mathrm{kpc} \le d_{\mathrm{MW-LMC}} \le 100 \, \mathrm{kpc}$ in the $\Lambda$CDM cosmological simulations TNG50-1 (46 objects; \emph{left}), TNG100-1 (118 objects; \emph{middle}), and TNG300-1 (191 objects; \emph{right}). The red cross shows the observed values, i.e. $f_{\mathrm{MCs,obs}} = 9.10 \times 10^{-11} \, \mathrm{km^{-3}\,s^{3}\,kpc^{-3}}$ and $\Omega_{\mathrm{LMC,obs}} = 6.48 \, \mathrm{km\,s^{-1}\,kpc^{-1}}$. None of the 355 analogues in the three TNG runs have $f_{\mathrm{MCs}} \ge f_{\mathrm{MCs,obs}}$ and $\Omega_{\mathrm{LMC}} \ge \Omega_{\mathrm{LMC,obs}}$, as indicated by the grey shaded region at the top right of each panel.}
    \label{figure_phase_space}
\end{figure*}

The distributions of $f_{\mathrm{MCs}}$ (Equation~\ref{eq:specific_phase_space_density}) and $\Omega_{\mathrm{LMC}}$ (Equation~\ref{eq:velocity_distance_LMC}) for these systems are shown in Figure~\ref{figure_phase_space}. In all three runs, none of the 355 selected MCs analogues have $f_{\mathrm{MCs}} \ge f_{\mathrm{MCs,obs}}$ and $\Omega_{\mathrm{LMC}} \ge \Omega_{\mathrm{LMC,obs}}$, as indicated by the grey region. This conclusion remains the same if we only consider the condition $f_{\mathrm{MCs}} \ge f_{\mathrm{MCs,obs}}$, indicating that the phase-space density of the MCs is the most problematic aspect of the observations. 3 out of the 355 systems satisfy $\Omega_{\mathrm{LMC}} \ge \Omega_{\mathrm{LMC,obs}}$.

In the high-resolution realization TNG50-1, the null detection of MCs analogues with $f_{\mathrm{MCs}} \ge f_{\mathrm{MCs,obs}}$ means that the frequency of MCs analogues is $<1/46$, which is equivalent to a tension of $>2.29\sigma$ for a single Gaussian variable (Equation~\ref{P_chi}). In TNG100-1 and TNG300-1, the upper limit to the frequency of such MCs analogues is $1/118$ and $1/191$, corresponding to a $>2.63\sigma$ and $>2.79\sigma$ tension, respectively. This gives the misleading impression that the tension decreases in the higher resolution runs. However, the box size, and therefore also the sample size of MCs analogues, is smaller in the higher resolution runs. As a consequence, the lower limit to the tension becomes minimal in TNG50-1, giving the misleading impression that this simulation is in agreement with $f_{\mathrm{MCs,obs}}$. 

\begin{table*}
  \centering
    \resizebox{\linewidth}{!}{\begin{tabular}{lllll}
    \hline
    Selection criteria & Simulation & Number of analogues & Frequency & Significance \\ \hline
    Initial sample of analogues to the MW (Section~\ref{subsec:Selection criteria}) & TNG50-1: & $1547$ & $-$ & $-$ \\
    & TNG100-1: & $7360$ & $-$ & $-$ \\
    & TNG300-1: & $40075$ & $-$ & $-$ \\  
    & All 3 runs: & $48982$ & $-$ & $-$ \\ \hline 
    Initial sample of analogues to the LMC (Section~\ref{subsec:Selection criteria of MCs-analogues}) & TNG50-1: & $331$ & $~331/1547$ & $~1.24 \sigma$  \\
    & TNG100-1: & $1414$ & $~1414/7360$ & $~1.30 \sigma$ \\
    & TNG300-1: & $3823$ & $~3823/40075$ & $~1.67 \sigma$ \\ 
    & All 3 runs: & $5568$ & $~5568/48982$ & $~1.58 \sigma$ \\ \hline 
    
    Initial sample of analogues to the SMC (Section~\ref{subsec:Selection criteria of MCs-analogues}) & TNG50-1: & $343$ & $~343/1547$ & $~1.22 \sigma$  \\
    & TNG100-1: & $766$ & $~766/7360$ & $~1.63 \sigma$ \\
    & TNG300-1: & $1148$ & $~1148/40075$ & $~2.19 \sigma$ \\ 
    & All 3 runs: & $2257$ & $~2257/48982$ & $~1.99 \sigma$ \\ \hline 
    
    Initial sample of analogues to the MCs (Section~\ref{subsec:Selection criteria of MCs-analogues}) & TNG50-1: & $147$ & $~147/1547$ & $~1.67 \sigma$  \\
    & TNG100-1: & $454$ & $~454/7360$ & $~1.87 \sigma$ \\
    & TNG300-1: & $601$ & $~601/40075$ & $~2.43 \sigma$ \\ 
    & All 3 runs: & $1202$ & $~1202/48982$ & $~2.25 \sigma$ \\ \hline
    
    Initial sample of analogues to the MCs (Section~\ref{subsec:Selection criteria of MCs-analogues}) & TNG50-1: & $46$ & $~46/1547$ & $~2.17 \sigma$  \\
    +$50 \le d_{\mathrm{MW-LMC}}/\mathrm{kpc} \le 100$& TNG100-1: & $118$ & $~118/7360$ & $~2.41 \sigma$ \\
    & TNG300-1: & $191$ & $~191/40075$ & $~2.82 \sigma$ \\ 
    & All 3 runs: & $355$ & $~355/48982$ & $~2.69 \sigma$ \\ \hline 

    Sample of analogues to the MCs with $50 \le d_{\mathrm{MW-LMC}}/\mathrm{kpc} \le 100$ & TNG50-1: & $0$ & $<1/46$ & $>2.29 \sigma$  \\
    $+f_{\mathrm{MCs}} \ge f_{\mathrm{MCs,obs}}$ (Equation~\ref{eq:specific_phase_space_density})& TNG100-1: & $0$ & $<1/118$ & $>2.63 \sigma$ \\
    & TNG300-1: & $0$ & $<1/191$ & $>2.79 \sigma$ \\ 
    & All 3 runs: & $0$ & $<1/355$ & $>2.99 \sigma$ \\ \hline 

    Sample of analogues to the MCs with $50 \le d_{\mathrm{MW-LMC}}/\mathrm{kpc} \le 100$ & TNG50-1: & $-$ & $1.62 \times 10^{-3}$ & $3.15 \sigma$  \\
    Extrapolation of $f_{\mathrm{MCs}}$ distribution using a linear fit (Figure~\ref{figure_phase_space_extrapolation})& TNG100-1: & $-$ & $1.35 \times 10^{-2}$ & $2.47 \sigma$ \\
    & TNG300-1: & $-$ & $1.32 \times 10^{-2}$ & $2.48 \sigma$ \\ \hline    
    
    Sample of analogues to the MCs with $50 \le d_{\mathrm{MW-LMC}}/\mathrm{kpc} \le 100$ & TNG50-1: & $-$ & $7.81 \times 10^{-5}$ & $3.95 \sigma$  \\
    Extrapolation of $f_{\mathrm{MCs}}$ distribution using a quadratic fit (Figure~\ref{figure_phase_space_extrapolation})& TNG100-1: & $-$ & $1.87 \times 10^{-3}$ & $3.11 \sigma$ \\
    & TNG300-1: & $-$ & $4.84 \times 10^{-3}$ & $2.82 \sigma$ \\ \hline 

    Sample of analogues to the MCs with $50 \le d_{\mathrm{MW-LMC}}/\mathrm{kpc} \le 100$ & TNG50-1: & $0$ & $<6.85 \times 10^{-3}$ & $>2.70 \sigma$  \\
    $+f_{\mathrm{MCs}} \ge f_{\mathrm{MCs,obs}}$& TNG100-1: & $0$ & $<2.98 \times 10^{-3}$ & $>2.97 \sigma$ \\
    +LMC-DoS alignment & TNG300-1: & $0$ & $<1.93 \times 10^{-3}$ & $>3.10 \sigma$ \\ 
    & All 3 runs: & $0$ & $<1.10 \times 10^{-3}$ & $>3.26 \sigma$ \\ \hline 
    
    Sample of analogues to the MCs with $50 \le d_{\mathrm{MW-LMC}}/\mathrm{kpc} \le 100$ & TNG50-1: & $-$ & $6.65 \times 10^{-4}$ & $3.40 \sigma$  \\
    Extrapolation of $f_{\mathrm{MCs}}$ distribution using a linear fit (Figure~\ref{figure_phase_space_extrapolation})& TNG100-1: & $-$ & $4.52 \times 10^{-3}$ & $2.84 \sigma$ \\
    +LMC-DoS alignment & TNG300-1: & $-$ & $4.43 \times 10^{-3}$ & $2.85 \sigma$ \\  \hline     
    
    Sample of analogues to the MCs with $50 \le d_{\mathrm{MW-LMC}}/\mathrm{kpc} \le 100$ & TNG50-1: & $-$ & $3.90 \times 10^{-5}$ & $4.11 \sigma$  \\
    Extrapolation of $f_{\mathrm{MCs}}$ distribution using a quadratic fit (Figure~\ref{figure_phase_space_extrapolation})& TNG100-1: & $-$ & $7.55 \times 10^{-4}$ & $3.37 \sigma$ \\
    +LMC-DoS alignment & TNG300-1: & $-$ & $1.80 \times 10^{-3}$ & $3.12 \sigma$ \\  \hline 
	\end{tabular}}
	\caption{Analogues to the MCs (i.e., systems with satellites analogous to both the LMC and the SMC) in the redshift range $0 \le z \le 0.26$ in the TNG50-1, TNG100-1, and TNG300-1 simulation for different selection criteria. The table lists the number of analogues, their frequency relative to the total number of selected MW-like galaxies hosting analogues to both MCs, and the corresponding equivalent number of standard deviations for a single Gaussian variable (Equation~\ref{P_chi}). The first part summarizes the initial samples of analogues to the MW, LMC, SMC, and both MCs as defined in Sections~\ref{subsec:Selection criteria} and \ref{subsec:Selection criteria of MCs-analogues}. The other parts list the results if additional selection criteria (indicated by the plus symbol in the first column) are applied to the initial sample of analogues to the MCs. The last part of the table gives the combined likelihood of the MCs by adding up the $\chi^{2}$ values of their phase-space density (Section~\ref{subsubsec:Phase-space density}) and the alignment of the LMC orbital pole with that of the DoS. The corresponding $p$-value is calculated for two degrees of freedom (Section~\ref{subsubsec:Alignment of the LMC with the DoS}). As explained in Section~\ref{subsubsec:Phase-space density} and Appendix~\ref{subsec:Comparing polynomial fits}, the linear extrapolation underestimates the tension because the fit does not account for the curvature of the data at $\log_{10} \left( f_{\mathrm{MCs,min}}/\mathrm{\left[ km^{-3}\,s^{3}\,kpc^{-3} \right]} \right) \ga \, -13$.}
  \label{tab:results_LMC_SMC_analogues}
\end{table*}

\begin{figure}
    \includegraphics[width=\columnwidth]{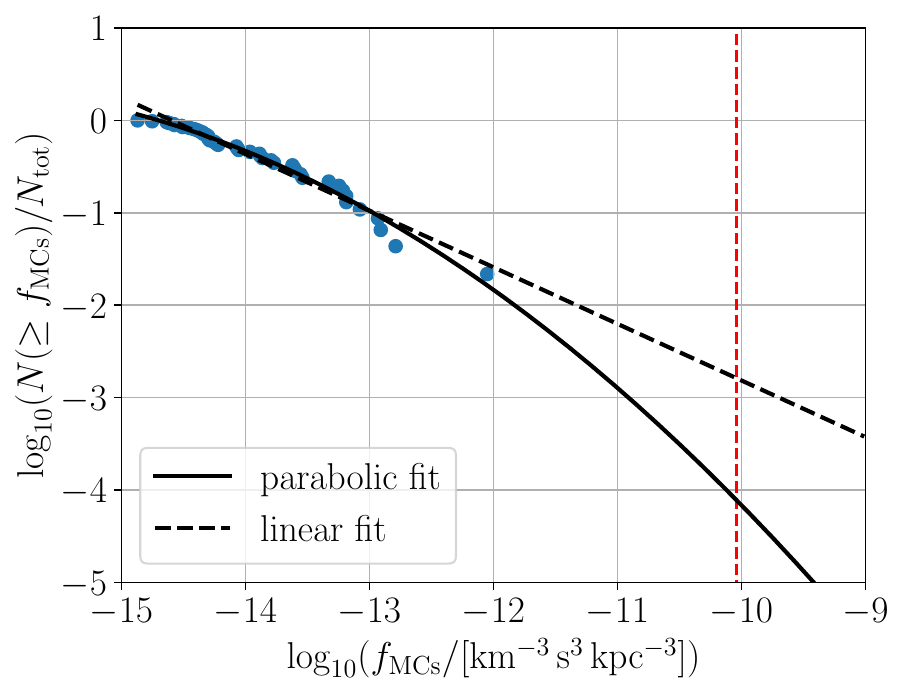}
    \caption{Cumulative distribution of $f_{\mathrm{MCs}}$ obtained from the TNG50-1 run fitted with a linear (dashed black line) and a quadratic (solid black line) function in $\log_{10}$-space. The red dashed line marks the observed value of $f_{\mathrm{MCs,obs}} = 9.10 \times 10^{-11} \, \mathrm{km^{-3}\,s^{3}\,kpc^{-3}}$. The intersection between the linear (quadratic) fit and the red dashed line yields a $p$-value of $1.62 \times 10^{-3}$ ($7.81 \times 10^{-5}$), which corresponds to a $3.15\sigma$ ($3.95\sigma$) tension for a single Gaussian variable (see also Table~\ref{tab:results_LMC_SMC_analogues} for TNG100-1 and TNG300-1). Polynomial fits up to order 3 and their residuals are shown in Appendix~\ref{subsec:Comparing polynomial fits}.}
    \label{figure_phase_space_extrapolation_TNG50-1}
\end{figure}

The above analysis only sets lower limits to the tension by relying on the null detection of MCs analogues in the different TNG runs. In the following, we estimate the significance by extrapolating the cumulative $f_{\mathrm{MCs}}$ distribution (i.e. the likelihood of an even higher value) up to the observed value $f_{\mathrm{MCs,obs}}$ \citep[see, e.g.,][]{Asencio_2021}. We fit the cumulative $f_{\mathrm{MCs}}$ distribution with a linear and a quadratic regression in $\log_{10}$-space and extrapolate the frequency of analogues up to $f_{\mathrm{MCs,obs}}$. This is visualized in Figure~\ref{figure_phase_space_extrapolation_TNG50-1} for the TNG50-1 run. The extrapolated cumulative frequency at $f_{\mathrm{MCs,obs}}$ is then converted into an equivalent number of standard deviations for a single Gaussian variable (Equation~\ref{P_chi}). Applying the linear (quadratic) regression to the entire range of the simulated $f_{\mathrm{MCs}}$ distribution results in a $3.15 \sigma$ ($3.95 \sigma$), $2.47\sigma$ ($3.11\sigma$), and $2.48 \sigma$ ($2.82\sigma$) tension in the TNG50-1, TNG100-1, and TNG300-1 run, respectively (see also Table~\ref{tab:results_LMC_SMC_analogues}). Thus, the tension becomes maximal in the highest-resolution realization TNG50-1. We will argue in Appendix~\ref{subsubsec:Effect of resolution} that the phase-space density configuration of simulated MCs analogues is likely not affected by numerical resolution issues in this simulation.

Figure~\ref{figure_phase_space_extrapolation_TNG50-1} indicates that the quadratic regression is a good fit especially for $\log_{10} \left( f_{\mathrm{MCs,min}}/\mathrm{\left[ km^{-3}\,s^{3}\,kpc^{-3} \right]} \right) \la \, -13$. The linear fit is more conservative with regards to the tension because it suppresses the curvature for data with $\log_{10} \left( f_{\mathrm{MCs,min}}/\mathrm{\left[ km^{-3}\,s^{3}\,kpc^{-3} \right]} \right) \ga \, -13$. However, this is inaccurate because curvature is apparent in the data. We quantify this using the Akaike Information Criterion \citep[AIC;][]{Akaike_1974} and the Bayesian Information Criterion \citep[BIC;][]{Schwarz_1978}, which are two of the most commonly used information criteria that have been proposed in the literature to quantify if adding an extra fit parameter improves the goodness of fit enough to justify the extra model complexity. This is a modern version of Occam's Razor. It can be phrased as quantifying an adjusted log-likelihood
\begin{eqnarray}
    \ln \widetilde{P} ~=~ \ln P - OP \, ,
\end{eqnarray}
where $\ln P$ is the likelihood of the best-fitting model matching the data, but this is then reduced by the `Occam penalty' $OP$, which can in principle be any function that takes a higher value for a more complicated model. In our case, we quantify $P$ by assuming Gaussian statistics:
\begin{eqnarray}
    P ~=~ \prod\limits_{i = 1}^N \frac{1}{\sigma \sqrt{2\mathrm{\pi}}} \exp \left[ -\frac{\left( y_i - y^{\mathrm{fit}}_i \right)^2}{2 \sigma^2} \right] \, ,
    \label{P_fit}
\end{eqnarray}
where $i$ labels the $N$ different data points $\left(x_i, y_i \right)$, the fit predicts a value $y^{\mathrm{fit}}_i$, and the scatter $\sigma$ about the fitted relation is estimated as
\begin{eqnarray}
    \sigma ~=~ \sqrt{\frac{\sum\limits_{i = 1}^N \left( y_i - y^{\mathrm{fit}}_i \right)^2}{N - f}} \, ,
\end{eqnarray}
where $f$ is the number of fit parameters, which is 2 for a linear fit and 3 for a quadratic fit. The use of $\left( N - f \right)$ rather than $N$ in the denominator captures the fact that a linear fit to 2 data points is guaranteed to exactly match both, so $\sigma$ would be biased low without this adjustment. Since $\sigma$ is estimated from the data, the exponential product in Equation~\ref{P_fit} is approximately the same regardless of the model, with the main difference being in $\sigma$ (roughly speaking, its estimated value is such that the total $\chi^2 = N - f$). Therefore, up to constant additive terms which have no bearing on the model selection problem, we can approximate that the best model minimises $N \ln \sigma + OP$.

Without the Occam penalty, we could always increase $P$ by having more parameters in our fit. But with the penalty term included, this is no longer guaranteed. To check whether a more complicated model is better, we merely need to check if $\ln \widetilde{P}$ increases, i.e. we need to check the difference $\Delta \ln \widetilde{P}$ between the two models. This arises from both $\Delta \ln P$ and $\Delta OP$, the increase in $OP$ associated with adding an extra model parameter.
\begin{eqnarray}
    \Delta OP = \begin{cases}
    1 & \textrm{(AIC)}\, , \\
    \frac{\ln N}{2} & \textrm{(BIC)} \, .
    \end{cases}
    \label{Delta_OP}
\end{eqnarray}

Applying these arguments to our case with $N = 46$ data points where the independent variable is $\log_{10} f_{\mathrm{MCs}}$ and the dependent variable is $\log_{10} N \left ( \ge f_{\mathrm{MCs}} \right)$, we get that $\ln P = 55.9$ for the linear fit and $\ln P = 69.5$ for the quadratic fit. We checked that these values remain the same if we instead maximise $\ln P$ by jointly varying $\sigma$ and the polynomial coefficients, all of which are almost exactly the same as with our analytic calculations based on standard polynomial regression. Both ways of viewing the problem show that the quadratic fit achieves a substantial improvement of $\Delta \ln P = 13.6$ relative to the linear fit. This is much greater than $\Delta OP$, which is only 1 with the AIC and 1.9 with the BIC. Therefore, it is clear that despite the higher complexity of a quadratic fit, it is indeed a significant improvement over a linear fit. This is also apparent from our results in Appendix~\ref{subsec:Comparing polynomial fits}, where we show the linear, quadratic, and cubic fits alongside the residuals in each case. Going from a quadratic to a cubic fit only slightly reduces $\sigma$ and is only mildly preferred by the AIC and BIC given that $\Delta \ln P = 4.2$, which barely exceeds $\Delta OP$. The extrapolated result with a cubic fit is intermediate between the results for the linear and quadratic fits, so these capture the plausible range of uncertainty in the likelihood of the observed $f_{\mathrm{MCs}}$. For simplicity, cubic and higher order polynomial fits are not considered further.

\begin{figure*}
    \includegraphics[width=\columnwidth]{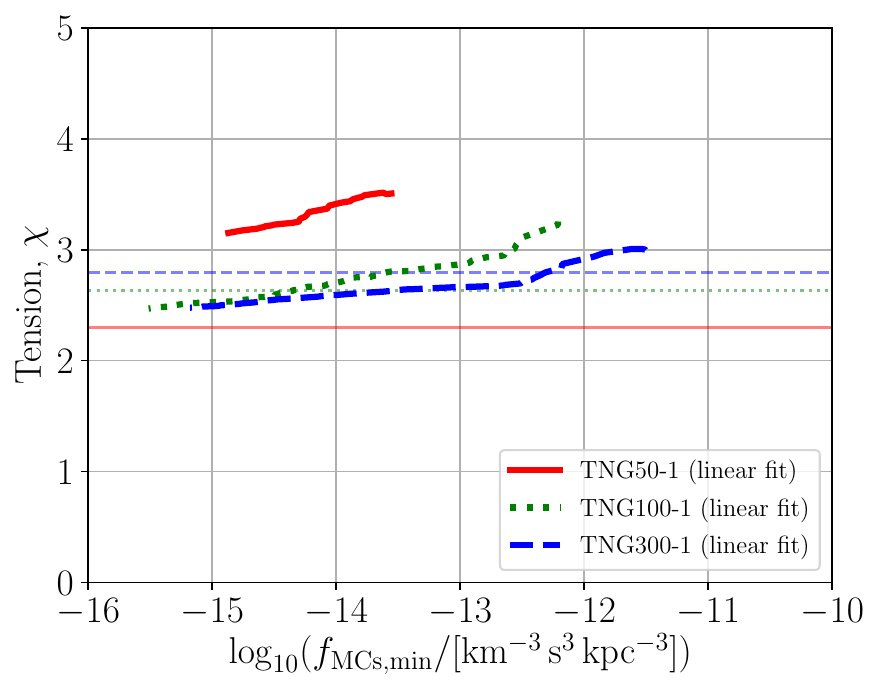}
    \includegraphics[width=\columnwidth]{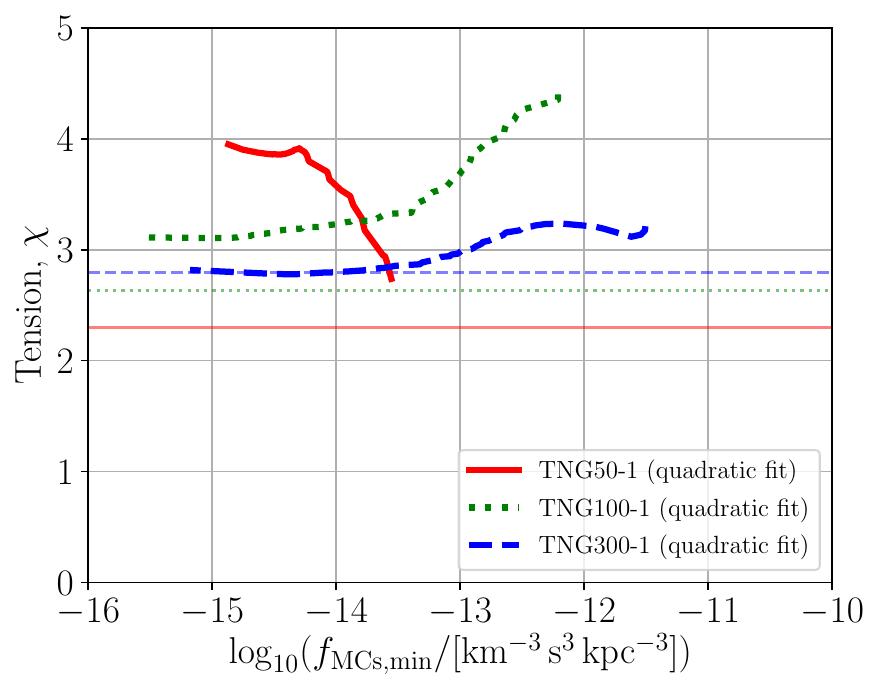}
    \caption{Estimated tension of the MCs' phase-space density in dependence of the minimum value of $f_{\mathrm{MCs}}$ beyond which we fit the cumulative $f_{\mathrm{MCs}}$ distribution of each simulation. The \emph{left} panel uses a linear fit and the \emph{right} panel uses a quadratic fit. The horizontal lines correspond to the null detection of analogues as equivalent to an upper limit on the frequency of analogues.}
    \label{figure_phase_space_extrapolation}
\end{figure*}

The frequencies and tensions for different selection criteria and statistical methods are summarized in Table~\ref{tab:results_LMC_SMC_analogues}. In order to assess the robustness of this method, we also derive the tension by fitting the distribution over different ranges of $f_{\mathrm{MCs}}$. For this, we fix the upper limit of the fitting range at the maximum value of $f_{\mathrm{MCs}}$ and perform different extrapolations by successively increasing the minimum limit of the fitting range, i.e. we only fit towards the upper end of the $f_{\mathrm{MCs}}$ distribution rather than using the full range. The so-estimated tensions in dependence of the minimum applied $f_{\mathrm{MCs}}$ limits are shown in Figure~\ref{figure_phase_space_extrapolation}. In TNG100-1 and TNG300-1, the tension converges to $\approx3.1\sigma$ and $\approx2.8\sigma$ for $\log_{10}\left(f_{\mathrm{MCs,min}}/\mathrm{\left[ km^{-3}\,s^{3}\,kpc^{-3}\right]}\right) \, \la \, -15$ and $\la \, -14$ for the quadratic fit (right panel), respectively. In TNG50-1, the tension reaches $\approx3.9\sigma$ for $\log_{10}\left(f_{\mathrm{MCs,min}}/\mathrm{\left[ km^{-3}\,s^{3}\,kpc^{-3}\right]}\right) \, \la \, -14.2$. In all cases, we only consider a fitting range wide enough to include at least ten points.

To assess the impact of different distance criteria, Appendix~\ref{subsec:Different selection criteria} discusses the $f_{\mathrm{MCs}}$ and $\Omega_{\mathrm{LMC}}$ distributions for analogues to the MCs with $d_{\mathrm{MW-LMC}} \ge 50$~kpc. None of the so-selected 1193 MCs analogues in the six TNG runs have $\Omega_{\mathrm{LMC}} \ge \Omega_{\mathrm{LMC,obs}}$ and $f_{\mathrm{MCs}} \ge f_{\mathrm{MCs,obs}}$, while only one of these analogues has $f_{\mathrm{MCs}} \ge f_{\mathrm{MCs,obs}}$. Possible reasons for the discrepancy between the observed phase-space density of the MCs and the $f_{\mathrm{MCs}}$ distribution of the TNG simulations are discussed further in Section~\ref{subsec:Effect of dynamical friction}.

\subsection{Tracing back analogues to the MCs in light of the MS}
\label{subsubsec:Individual analogues}

Using the initial sample of MCs analogues (Section~\ref{subsec:Selection criteria of MCs-analogues}), we found that 3 out of the 1202 analogues in TNG50-1, TNG100-1, and TNG300-1 combined fulfil $f_{\mathrm{MCs}} \ge f_{\mathrm{MCs,obs}}$ (Appendix~\ref{subsec:Different selection criteria}). In order to understand their formation and evolution in a cosmological context, we trace these analogues back through cosmic time using the \textsc{sublink} merger tree catalogue \citep{RodriguezGomez_2015}. We already know that in these systems, the LMC will be either very close to the MW (within 50~kpc) or very far away (beyond 100~kpc), making them somewhat different to the observed MCs. We nevertheless consider these systems in more detail to better understand the history behind $\Lambda$CDM systems that might come close to resembling the observed MCs.

\begin{figure*}
    \includegraphics[width=\columnwidth]{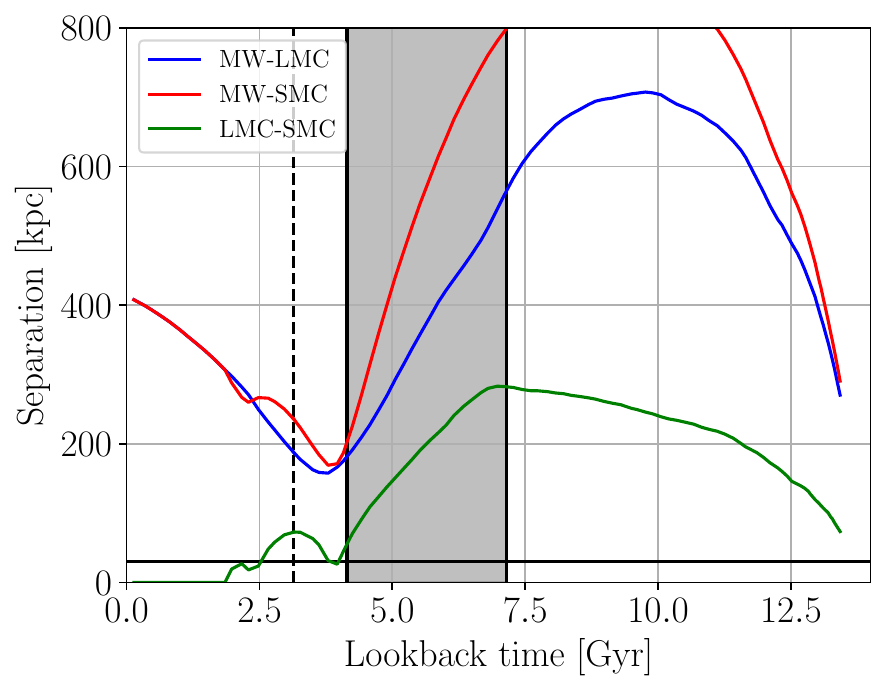}
    \includegraphics[width=\columnwidth]{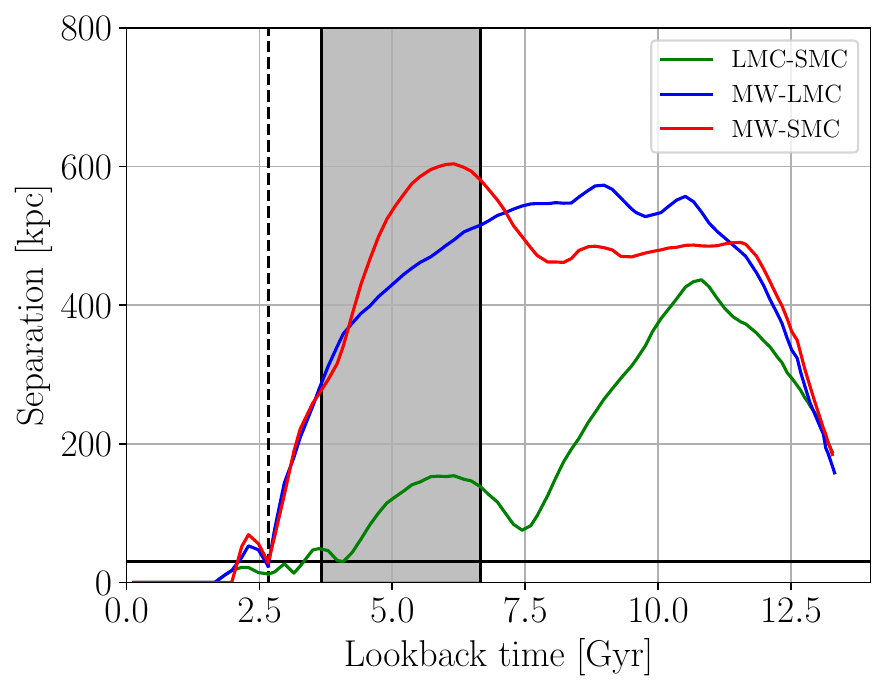}
    \caption{Time evolution of the physical separation between the MW-LMC (blue), MW-SMC (red), and LMC-SMC (green) in analogue system 1 (\emph{left} panel) and 2 (\emph{right} panel), which are identified at a lookback time of $3.1\,\mathrm{Gyr}$ ($z = 0.26$) in the TNG100-1 simulation and $2.7 \, \mathrm{Gyr}$ ($z = 0.21$) in the TNG300-1 simulation, respectively. Systems 2 and 3 consist of the same system identified as an analogue to the MCs at slightly different times (the larger lookback time is plotted here as the former would overlap with the 2.5~Gyr gridline). The formation of the MS requires a close interaction between the MCs before identification (see the text), for which we show a solid black horizontal line at $30\,\mathrm{kpc}$ and shade the time period $1-4$~Gyr before identification (dashed vertical line). \emph{Left panel:} Since there is no close ($<30$~kpc) interaction within this period, this system would not form an MS-like structure. \emph{Right panel:} The LMC and SMC have a pericentre passage wrt. each other about 1.5~Gyr before identification with $d_{\mathrm{LMC-SMC}} = 29.9 \, \mathrm{kpc}$, in principle marginally allowing the formation of an MS-like structure. The physical properties of these galaxies are listed in Table~\ref{tab:results_analogues_full_table}.}
    \label{figure_results}
\end{figure*}

The evolution of the physical separation between the MW-LMC, MW-SMC, and LMC-SMC analogues are shown in Figure~\ref{figure_results}. The first of these three analogues (hereafter system 1) is identified in the TNG100-1 simulation at redshift $z = 0.26$, which corresponds to a lookback time of 3.0~Gyr. In what follows, we will assume that the time of identification of a system would be the present time if we were living in that system's MW analogue. We will therefore use `$x$~Gyr ago' to mean $x$~Gyr prior to the time of identification. Both the MW-LMC and MW-SMC orbits have their first pericentre passage 0.6~Gyr ago, when $d_{\mathrm{MW-LMC}} = 157.6\,\mathrm{kpc}$ and $d_{\mathrm{MW-SMC}} = 169.0 \, \mathrm{kpc}$. As mentioned above, these greatly exceed the present Galactocentric distances of the MCs. The LMC and SMC analogues are accreted together towards the MW-like galaxy and have their first pericentre passage wrt. each other 0.8~Gyr ago, when $d_{\mathrm{LMC-SMC}} = 26.3\,\mathrm{kpc}$. At the time of identification, the MCs are at apocentre with $d_{\mathrm{LMC-SMC}} = 72.5\,\mathrm{kpc}$. The LMC has to tidally remove gas from the SMC during a past close interaction $\approx 3\,\mathrm{Gyr}$ ago to allow the stripped gas to almost circumnavigate the Galaxy, which is necessary to form the MS \citep{Donghia_2016}. In the model of \citet{Lucchini_2020}, a substantial part of the MS is formed during the second close LMC-SMC interaction about 1.5~Gyr ago at a pericentre separation of about 20~kpc (see their extended data figure~2). An interaction between the MCs $\ga 1$~Gyr ago could also be required to explain the northern arm of the LMC and substructures in the LMC's outskirts \citep{Cullinane_2022a, Cullinane_2022b}. Therefore, we consider it possible that the MS would form in our selected analogues only if the LMC and SMC have undergone a pericentre passage within 30~kpc over the time period $1-4$~Gyr before identification. We defined these criteria before extracting the trajectories of analogues in order to guarantee an unbiased evaluation of the situation. Such a close previous interaction between the MCs is absent in the here identified system 1, implying that it is inconsistent with the observed MS. While there is a close interaction within 30~kpc, we argue that 0.8~Gyr is too little time for the stripped gas to circumnavigate the galaxy, especially since the LMC and SMC in this system were rather widely separated at earlier times. Tracing the individual galaxies forward in cosmic time reveals that these analogue MCs merge with each other 1.3~Gyr after identification. Such short merger times are expected due to dynamical friction between the extended dark matter halos, as discussed further in Section~\ref{subsec:Effect of dynamical friction}.

%This is followed by a second pericentre $6.7 \,\mathrm{Gyr}$ ago with $d_{\mathrm{LMC-SMC}} = 57.4 \,\mathrm{kpc}$. The third pericentre passage occurs at the time of identification with a separation of $d_{\mathrm{LMC-SMC}} = 22.5 \,\mathrm{kpc}$.  %This is particularly apparent in the top right panel of Figure~\ref{figure_results}, which provides a zoomed-in view. 

The second and third analogues correspond to the same system identified at different times in TNG300-1. The formation and evolution of this system are presented in the right panel of Figure~\ref{figure_results}. This system satisfies $f_{\mathrm{MCs}} \ge f_{\mathrm{MCs,obs}}$ at $z = 0.21$ (system~2) and $z = 0.20$ (system~3), corresponding to lookback times of 2.7~Gyr and 2.5~Gyr, respectively. Similarly to system 1 and using the $z = 0.21$ case, the LMC and SMC analogues are accreted together towards the MW-like galaxy. When the system fulfils the selection criteria at $z = 0.21$, the MW-LMC and MW-SMC have their first pericentre but the LMC-SMC system has its fourth pericentre: the first occurred $4.8 \,\mathrm{Gyr}$ ago ($d_{\mathrm{LMC-SMC}} = 75.3 \,\mathrm{kpc}$), the second was 1.4~Gyr ago ($d_{\mathrm{LMC-SMC}} = 29.9 \,\mathrm{kpc}$), and the third was 0.5~Gyr ago ($d_{\mathrm{LMC-SMC}} = 13.3 \,\mathrm{kpc}$). Thus, in contrast to the previous example, the close interaction of the LMC-SMC system before its detection (grey shaded region in Figure~\ref{figure_results}) would in principle allow the formation of a gaseous tidal stream comparable to the MS. Note that this system only marginally satisfies our MS condition because there is only one pericentre $1-4$~Gyr before identification and the distance then is 29.9~kpc, which only just satisfies our requirement to be within 30~kpc. The hydrodynamical simulations of \citet{Lucchini_2021} assume a much closer interaction at a separation well below 20 kpc, as shown in their figure~1b. It would be interesting to check to what extent a more distant interaction between the MCs might still be able to form the MS.

Our results support a first infall scenario in which the MCs are accreted together within the last $\la 3 \, \mathrm{Gyr}$, in agreement with previous studies \citep[e.g.,][]{BoylanKolchin_2011}. This causes a high relative velocity of the LMC analogue wrt. the MW-like galaxy and a high relative velocity between the MCs of $173.3 \, \mathrm{km\,s^{-1}}$ at $z=0.21$ (system 2) and $148.0 \, \mathrm{km\,s^{-1}}$ at $z = 0.20$ (system 3), which significantly exceeds the observed $v_{\mathrm{LMC-SMC}} = 90.8 \, \mathrm{km\,s^{-1}}$ in all cases. Furthermore, the MW-LMC and MW-SMC orbital poles are misaligned by $59^\circ.65$ at $z = 0.21$ (system 2) and by $115^\circ.67$ at $z=0.20$ (system 3), being therewith much less aligned than observed ($\theta = 16^\circ.96$). Note that this MW-MCs system is again short-lived as the LMC (SMC) merges with the MW analogue 1.0~Gyr (0.7~Gyr) after identification.

The physical properties of all three MW-MCs analogues are listed in Table~\ref{tab:results_analogues_full_table}.

\renewcommand{\arraystretch}{1.2}
\begin{table*}
  \centering
    \begin{tabular}{lllllll}
    \hline
    Parameter & Unit & Observed & System 1 & System 2 & System 3 \\ \hline
    Simulation & $-$ & $-$ & TNG100-1 & TNG300-1 & TNG300-1 \\ 
    Redshift & $-$ & $\approx 0$ & 0.26 & 0.21 & 0.20 \\ 
    Lookback time & Gyr & $\approx 0$ & $3.1$ & $2.7$ & $2.5$ \\ 
    $M_{\mathrm{total}}^{\mathrm{MW}}$ & $10^{10} \, M_{\odot}$ & $-$ & $171$ & $232$ & $232$ \\
    $M_{\mathrm{dm}}^{\mathrm{MW}}$ & $10^{10} \,M_{\odot}$ & $-$ & $159$ & $202$ & $202$ \\
    $M_{\star}^{\mathrm{MW}}$ & $10^{10} \, M_{\odot}$  & $(6.08 \pm 1.14)$ & $5.12$ & $5.14$ & $5.51$\\
    $M_{\mathrm{total}}^{\mathrm{LMC}}$ & $10^{10} \, M_{\odot}$ & $-$ & $17.50$ & $5.51$ & $4.92$\\
    $M_{\mathrm{dm}}^{\mathrm{LMC}}$ & $10^{10} \,M_{\odot}$ &  $-$  & $15.34$ & $3.45$ & $2.98$ \\
    $M_{\star}^{\mathrm{LMC}}$ & $10^{10} \,M_{\odot}$ & $0.15$ &  $0.79$ & $0.96$ & $0.98$ \\
    $M_{\mathrm{total}}^{\mathrm{SMC}}$ & $10^{10} \,M_{\odot}$ & $-$ & $1.62$ & $0.48$ & $0.31$ \\
    $M_{\mathrm{dm}}^{\mathrm{SMC}}$ & $10^{10} \,M_{\odot}$ & $-$ & $1.45$ & $0.38$ & $0.25$ \\
    $M_{\star}^{\mathrm{SMC}}$ & $10^{10} \,M_{\odot}$ & $0.046$ & $0.051$ & $0.095$ & $0.060$ \\
    $d_{\mathrm{MW-LMC}}$ &$\mathrm{kpc}$ & 50.0  & 187.2 & 23.7 & 47.0 \\
    $d_{\mathrm{MW-SMC}}$ &$\mathrm{kpc}$ & 61.3 & 235.3 & 28.7 & 55.8 \\
    $d_{\mathrm{LMC-SMC}}$ &$\mathrm{kpc}$ & 24.5  & 72.5 & 11.6 & 14.03 \\
    $v_{\mathrm{MW-LMC}}$ & $\mathrm{km\,s^{-1}}$ & 323.8  & 213.2 & 381.2 & 92.3 \\
    $v_{\mathrm{MW-SMC}}$ & $\mathrm{km\,s^{-1}}$ & 245.6  & 192.2 & 248.3 & 185.7 \\
    $v_{\mathrm{LMC-SMC}}$ & $\mathrm{km\,s^{-1}}$ & 90.8  & 21.9 & 173.3 & 148.0 \\
    $h_{\mathrm{MW-LMC}}$ & $\mathrm{10^{3}\, kpc\,km\,s^{-1}}$ & 15.9  & 36.8 & 5.6 & 2.1 \\
    $h_{\mathrm{MW-SMC}}$ & $\mathrm{10^{3}\, kpc\,km\,s^{-1}}$ & 15.0  & 40.5 & 1.2 & 5.2 \\
    $f_{\mathrm{MCs}}$ (Equation~\ref{eq:specific_phase_space_density}) & $10^{-10}\,\mathrm{km^{-3}\,s^{3}\,kpc^{-3}}$ & 0.91 & 2.51 & 1.22 & 1.12 \\
    $\Omega_{\mathrm{LMC}}$ (Equation~\ref{eq:velocity_distance_LMC}) & $\mathrm{km\,s^{-1}\,kpc^{-1}}$ & 6.48 & 1.14 & 16.09 & 1.96 \\ 
    $\mathrm{arccos}(\widehat{\bm{h}}_{\mathrm{MW-LMC}} \cdot \widehat{\bm{h}}_{\mathrm{MW-SMC}} )$ & $-$ & $16^\circ.96$ & $15^\circ.66$ & $59^\circ.65$ & $115^\circ.67$ \\ 
    $\mathrm{arccos}(\widehat{\bm{h}}_{\mathrm{MW-LMC}} \cdot \widehat{\bm{h}}_{\mathrm{LMC-SMC}} )$ & $-$ & $132^\circ.68$ & $119^\circ.69$ & $87^\circ.59$ & $96^\circ.43$ \\ \hline 
	\end{tabular}
	\caption{Physical properties of the observed MW, LMC, and SMC (third column) and their analogues (later columns), which fulfil $f_{\mathrm{MCs}} \ge f_{\mathrm{MCs,obs}} = 9.1 \times 10^{-11} \, \mathrm{km^{-3}\,s^{3}\,kpc^{-3}}$ at redshift $z = 0.26$ (system~1), $z = 0.21$ (system~2), and $z = 0.20$ (system~3). Systems~2 and 3 consist of the same objects identified at different timesteps (snapshots). The last two rows list the alignment of the orbital poles, where $\widehat{\bm{h}}$ is the normalized specific angular momentum vector. The observed stellar masses of the MW and the MCs are taken from \citet{Licquia_2015} and \citet{McConnachie_2012}, respectively. The distances and velocities of the MCs are taken from \citet{Pawlowski_2020}. The dark matter and total masses of these galaxies are not listed because these parameters are very sensitive to the applied measurement method (though see Section~\ref{subsubsec:Total mass of LMC analogues}).}
  \label{tab:results_analogues_full_table}
\end{table*}
\renewcommand{\arraystretch}{1}

\section{Discussion}
\label{sec:Discussion}

In this section, we first comment on how the low $p$-values of our analysis compare to prior studies on the MCs in $\Lambda$CDM simulations (Section~\ref{subsec:Prior studies on MCs in LCDM}). Secondly, we discuss which physical processes might be responsible for the low frequencies of analogues to the MCs, paying particular attention to the effect of dynamical friction on their orbits (Section~\ref{subsec:Effect of dynamical friction}). Finally, we consider the relation between the MCs and the LG satellite planes (Section~\ref{subsec:MCs and VPoS}) and the dynamical mass of the LMC based on nearby tidal streams (Section~\ref{subsubsec:Total mass of LMC analogues}).

\subsection{Prior studies on the MCs in \texorpdfstring{$\Lambda$}{L}CDM}
\label{subsec:Prior studies on MCs in LCDM}

The dynamics and occurrence rate of analogues to the MCs have been considered in several previous studies in the $\Lambda$CDM framework \citep[e.g.,][]{BoylanKolchin_2011, Busha_2011, SantosSantos_2021}. In particular, \citet{BoylanKolchin_2011} examined the pure $N$-body simulation Millennium-II \citep[MS-II;][]{BoylanKolchin_2009}, finding that about 35\% (32\%) of MW-like halos with a virial mass in the range $10^{12} < M_{\mathrm{vir}}/M_{\odot} < 3 \times 10^{12}$ host an analogue to the LMC (SMC) at $z=0$. They defined an LMC (SMC) analogue satellite as that with the highest (second-highest) dark matter-only infall mass $M_{\mathrm{acc}}$, which furthermore needs to be in the range $8 \times 10^{10} < M_{\mathrm{acc}}/M_{\odot} < 3.2 \times 10^{11}$ for the LMC and $4 \times 10^{10} < M_{\mathrm{acc}}/M_{\odot} < 1.6 \times 10^{11}$ for the SMC \citep[see section~2.3.2 of][]{BoylanKolchin_2011}. Their frequency of LMC (35\%) and SMC (32\%) analogues in MW-like halos is higher than in our analysis of the TNG simulations, in which only 11.4\% and 4.6\% of the selected MW-like galaxies host at least one LMC and SMC analogue, respectively, if we combine results from TNG50-1, TNG100-1, and TNG300-1 (Table~\ref{tab:results_LMC_SMC_analogues}). One reason for the difference between their result and ours could be that we select analogues based on the stellar mass and $M_{\mathrm{total}}/M_{\star}$ (Section~\ref{subsec:Selection criteria of MCs-analogues}). MS-II is a purely $N$-body simulation (it only considers dark matter) whereas the here-assessed IllustrisTNG runs are hydrodynamical cosmological simulations, allowing us to select satellite galaxies based on their stellar mass without the need of semi-analytic models and abundance matching techniques. Since the stellar mass is directly observed, our comparison with observations should be more direct. 

Similarly to the three identified analogues to the MCs in Section~\ref{subsubsec:Individual analogues}, the LMC analogues found by \citet{BoylanKolchin_2011} were typically accreted at late times ($\la 4$~Gyr ago). About 2.5\% of their MW-like halos host LMC-SMC binaries with separation $d_{\mathrm{LMC-SMC}} < 50\,h^{-1}\,\mathrm{kpc} \approx 68.5\,\mathrm{kpc}$ and relative velocity $v_{\mathrm{LMC-SMC}} < 150 \, \mathrm{km\,s^{-1}}$ (see their section~5), which is comparable to the \emph{HST} three-epoch proper motion measurement \citep{Kallivayalil_2013}. These upper limits to the LMC-SMC separation and relative velocity are significantly higher than the observed values derived from the latest data ($d_{\mathrm{LMC-SMC}} = 24.5\,\mathrm{kpc}$ and $v_{\mathrm{LMC-SMC}} = 90.8 \, \mathrm{km\,s^{-1}}$). In particular, our assumed proper motions \citep[from][]{Pawlowski_2020} are based on a combination of results from \emph{Gaia}~DR2 \citep{Gaia_2018b} and the \emph{HST} \citep{Kallivayalil_2013}, both of which were published after the work of \citet{BoylanKolchin_2011}. Proper motion uncertainties generally inflate the estimated $v_{\mathrm{LMC-SMC}}$, so this may decrease further as observations improve. Furthermore, their section~5 defined the SMC analogue of an LMC-SMC binary system as the second-ranked subhalo in the host without setting a specific mass range, which could make the SMC analogue a lot less massive than the actual SMC. We thus expect their work to overestimate the frequency of analogues to the MCs.

In the Bolshoi simulation \citep{Klypin_2011, TrujilloGomez_2011}, about 10\% of hosts with a virial mass similar to the MW have two MC-like satellites based on the maximum circular velocity \citep[see section~2 and figure~1 of][]{Busha_2011_statistics}. In contrast to our work, those authors did not apply separation or velocity criteria between the MCs and the MW or between the MCs themselves. They subsequently used instead the $r$-band luminosity for a better comparison with observations (see their section~4 and table~2). Their results are very consistent with the observational findings of \citet{Liu_2011}, who found that $3.5\%$ of MW-like galaxies host two satellites with luminosities comparable to the MCs within a radius of 150~kpc based on the Sloan Digital Sky Survey \citep{SDSS}.

In the IllustrisTNG simulations, 9.5\% (TNG50-1), 6.2\% (TNG100-1), and 1.5\% (TNG300-1) of MW-like galaxies host two satellites with stellar masses similar to the MCs (Table~\ref{tab:results_LMC_SMC_analogues}). Thus, the highest resolution realization TNG50-1 is consistent with \citet{Busha_2011_statistics}. The frequency of analogues to the MCs is $1.5\times$ ($6.3\times$) higher in TNG50-1 compared to TNG100-1 (TNG300-1), which could be a resolution effect. However, this should not significantly affect our results in Section~\ref{sec:Results} because the phase-space density of analogues to the MCs seems to have converged in the IllustrisTNG runs: the $f_{\mathrm{MCs}}$ distribution peaks at $\log_{10}( f_{\mathrm{MCs}}/\left[\mathrm{km^{-3}\,s^{3}\,kpc^{-3}}\right]) \approx -14$ in all six analysed resolution realizations, though the dispersion is larger in the TNG100 and TNG300 runs because of the larger sample sizes (see Appendix~\ref{subsubsec:Effect of resolution}).

\subsection{First infall scenario}
\label{subsec:Effect of dynamical friction}

The observed LMC and SMC lie at Galactocentric distances of $50.0\,\mathrm{kpc}$ and $61.3\,\mathrm{kpc}$, respectively, with a mutual separation of only 24.5~kpc. This is much smaller than the virial radius of the MW, indicating that Chandrasekhar dynamical friction must be very efficient if the $\Lambda$CDM framework is correct because it predicts that primordial galaxies are embedded in extended CDM halos. If a satellite galaxy moves through such a halo, it loses kinetic energy and momentum via encounters with the CDM particles. 

The action of Chandrasekhar dynamical friction between dark matter halos limits the accessible phase-space number density and mass density. For example, it would be unlikely to observe three massive galaxies in close ($< 100\,\mathrm{kpc}$) proximity with small relative velocities ($< 100 \, \mathrm{km\,s^{-1}}$) due to the rapid merging timescale. However, just such a scenario arises with the MCs: their mutual separation is only $d_{\mathrm{LMC-SMC}} = 24.5 \, \mathrm{kpc}$, which combined with a low relative velocity of $v_{\mathrm{LMC-SMC}} = 90.8 \, \mathrm{km\,s^{-1}}$ implies a high phase-space density of $f_{\mathrm{MCs,obs}} = 9.10 \times 10^{-11} \, \mathrm{km^{-3}\,s^{3}\,kpc^{-3}}$. We showed that the phase-space density of the MCs is in $\approx 3 \sigma$ tension based on the TNG simulations (Section~\ref{subsubsec:Phase-space density}). The $f_{\mathrm{MCs}}$ distribution is similar for different resolution realizations, implying numerical robustness of our results (Appendix~\ref{subsubsec:Effect of resolution}).

The small expected phase-space density of the MCs is related to the first infall scenario, the most likely formation process of the MCs within the $\Lambda$CDM framework (Section~\ref{subsubsec:Individual analogues}). In this scenario, the MCs have fallen into the MW from large distances, which typically yields high relative velocities between the MCs and makes it unlikely that they are very close to each other. The identified analogue system 1 in Section~\ref{subsubsec:Individual analogues} nevertheless has $v_{\mathrm{LMC-SMC}} = 21.9 \,\mathrm{km\,s^{-1}}$, being therewith much smaller than the observed value. However, the MCs in this system have large Galactocentric distances of $d_{\mathrm{MW-LMC}} = 187.2$~kpc and $d_{\mathrm{MW-SMC}} = 235.3$~kpc (Table~\ref{tab:results_analogues_full_table}). In contrast, systems~2 and 3 have the MCs very close to the MW analogue. In these cases, the relative velocity between the MCs is somewhat higher than observed ($173.3\,\mathrm{km\,s^{-1}}$ in system~2 and $148.0\,\mathrm{km\,s^{-1}}$ in system~3). The low LMC-SMC relative velocity and the proximity of the MCs to the MW therefore argues against a scenario where both MCs fell in from large distances and only encountered each other in the last few~Gyr.

\subsection{Relating the MCs to the DoS} 
\label{subsec:MCs and VPoS}

The MCs are part of the DoS, a flattened and kinematically coherent plane perpendicular to the Galactic disc containing 7 or 8 of the 11 classical satellites \citep{Kroupa_2005, Metz_2007, Pawlowski_2012a, Pawlowski_2020}. While it is not our intention here to assess whether such a DoS is likely to arise in $\Lambda$CDM, we test the scenario that group infall of a massive galaxy like the LMC with its own retinue of satellites could explain the DoS (Section~\ref{subsubsec:Satellites of the LMC}). Since we will find that this is not a viable explanation, the DoS should have formed independently of the MCs. We therefore quantify how likely it is that the LMC would fall into a pre-existing DoS, and how this further worsens the situation for $\Lambda$CDM when considered together with the high phase-space density of the MCs (Section~\ref{subsubsec:Alignment of the LMC with the DoS}).

\subsubsection{Satellites of the LMC}
\label{subsubsec:Satellites of the LMC}

\begin{figure}
    \includegraphics[width=\columnwidth]{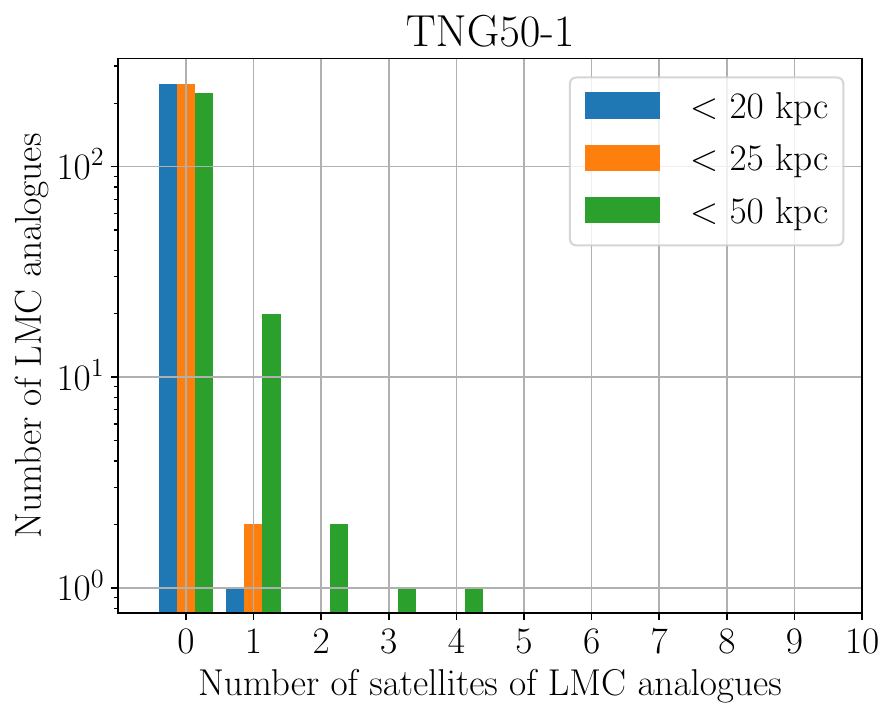} 
    \caption{Distribution of the number of subhalos with $M_{\star} > 10^{5} \, M_\odot$ within $20\,\mathrm{kpc}$ (blue), $25\,\mathrm{kpc}$ (orange), or $50\,\mathrm{kpc}$ (green) of pre-infall LMC analogues with $d_{\mathrm{MW-LMC}} > 100\,\mathrm{kpc}$ in the initial LMC sample (Section~\ref{subsec:Selection criteria of MCs-analogues}) in the high-resolution realization TNG50-1 ($248$ objects).}
    \label{figure_LMC_satellite}
\end{figure}

The group infall scenario states that dwarfs orbiting the LMC became classical MW satellites. Despite \citet{Metz_2009} showing that group infall cannot explain the DoS because known groups of dwarf galaxies have too large a spatial extent, it has nevertheless been argued that such an infall of satellite galaxies in groups can potentially explain the Galactic DoS \citep[e.g.,][]{DOnghia_2008, Li_2008_infall, Samuel_2021}. To estimate the number of satellites around an infalling LMC, we select galaxies with $d_{\mathrm{MW-LMC}} > 100$~kpc from the initial LMC sample of Section~\ref{subsec:Selection criteria of MCs-analogues}. The distribution of the number of subhalos with $M_{\star} > 10^{5}\,M_\odot$ within $20\,\mathrm{kpc}$, $25\,\mathrm{kpc}$, or $50\,\mathrm{kpc}$ of the infalling LMC analogue for the highest resolution realization TNG50-1 is shown in Figure~\ref{figure_LMC_satellite}. We select subhalos with any SubhaloFlag parameter (Section~\ref{subsec:Selection criteria of MCs-analogues}) to include subhalos with a non-cosmological origin, e.g. tidal dwarf galaxies (TDGs), disc structures in the host galaxy, etc. This is to be more conservative by overestimating the number of satellites around an infalling LMC.

The 11 classical satellites of the MW form a plane with a root mean square (rms) height of only $\Delta_{\mathrm{rms}} = 19.6\,\mathrm{kpc}$ \citep[table~1 of][]{Pawlowski_2021}, suggesting that 10 satellites should be distributed within a radius of $20\,\mathrm{kpc}$ around the infalling LMC. However, according to Figure~\ref{figure_LMC_satellite}, not a single infalling LMC analogue has more than one such satellite within $20$~kpc (or even $25$~kpc), even though 248 LMC analogues were considered altogether.

The resolution of the TNG100-1 and TNG300-1 (Section~\ref{subsec:Cosmological simulation}) runs is too low to evaluate the group infall scenario for the MW satellite galaxies. But TNG50-1 has an initial baryonic mass resolution of $8.5 \times 10^{4} \, M_\odot$, which is lower than the least massive classical satellites \citep[Draco and Ursa Minor have $M_{*} = 2.9 \times 10^{5} \, M_\odot$;][]{McConnachie_2012}. Such galaxies would only consist of a few stellar particles in the TNG50-1 run, implying they would not be resolved $-$ but most of the classical satellites should be resolved. It would be valuable to repeat the analysis using higher resolution realizations.

Relaxing the maximum allowed distance from the LMC to 50 kpc yields one LMC analogue with three satellites and a further one with four satellites. These satellites have stellar masses of $3.3 \times 10^{5}\,M_\odot$, $1.8 \times 10^{6}\,M_\odot$, and $2.2 \times 10^{7}\,M_\odot$ in the first case and $1.5 \times 10^{7}\,M_\odot$, $3.5 \times 10^{7}\,M_\odot$, $2.8 \times 10^{8}\,M_\odot$, and $1.4 \times 10^{8}\,M_\odot$ in the second case. These values are quite consistent with the stellar mass range of the 11 observed classical satellites of the MW \citep{McConnachie_2012}. However, even these two systems are unlikely to properly explain the very thin DoS because there are insufficiently many satellites around the LMC analogue and even these satellites are too far from it. More generally, it would be unusual if the LMC did bring in most of the classical satellites because this would imply that the MW had very few satellites of its own only a few Gyr ago, which seems rather unlikely in the $\Lambda$CDM paradigm because an MW-mass galaxy should have many more satellites than an LMC-mass galaxy.

Our findings are consistent with the results of \citet{SantosSantos_2021}, who found that about 2 satellites with $M_{\star} > 10^{5}\,M_\odot$ have been brought in with the LMC based on the APOSTLE simulations \citep{Fattahi_2016, Sawala_2016}, which have a similar resolution to TNG50-1 but focus on resimulating analogues to the LG in terms of environment and kinematics. \citet{GaravitoCamargo_2021} argued that the infall of an LMC-like galaxy yields orbital poles of dark matter particles aligned with the LMC orbital pole. They proposed that such an infall would similarly change the orbital poles of the other classical satellites, perhaps explaining the observed alignment of orbital poles in the DoS. However, \citet{Pawlowski_2022_LMC} showed that this effect would be too small to explain the high orbital pole density of the DoS once we consider that the satellites in it have high specific angular momenta, preventing the modest tidal torques from the LMC from significantly reorienting their orbital poles \citep[see also][]{Correa_2022}.

These problems have led to several studies on whether the DoS can be understood within $\Lambda$CDM, and what the latest data can tell us. It has been argued by \citet{Sawala_2023} that the orbital alignment of satellites in the DoS is less pronounced $-$ and thus less unusual for $\Lambda$CDM $-$ if we update the proper motions from \emph{Gaia}~DR2 to the latest \emph{Gaia}~DR3 \citep{Gaia_2021, Gaia_2023}. Unfortunately, this claim stems from an invalid comparison between \emph{Gaia}~DR3 without the \emph{HST} and an earlier result that combined \emph{Gaia}~DR2 with the \emph{HST} \citep{Pawlowski_2020}. Their table~3 shows that if the \emph{HST} results are not considered at all and we only use \emph{Gaia} data, the orbital pole dispersion of the seven most concentrated orbital poles is actually tighter with the latest \emph{Gaia} data according to the orbital pole calculations of \citet{Sawala_2023}. Therefore, the latest \emph{Gaia} data makes the DoS problem worse for $\Lambda$CDM.

As pointed out in section~5.1.3 of \citet{Kroupa_2015}, the recent infall of a group is inconsistent with the observed low gas content of most LG satellite galaxies. Indeed, \citet{Nichols_2011} estimated that the accretion of primordial satellites must have happened at much higher redshifts of $z = 3 - 10$ in order to account for most of the LG satellites being gas-depleted, therewith ruling out a recent accretion scenario. However, in order to retain the presently observed anisotropic satellite distribution, the infall of the group must have occurred at low redshifts \citep{Klimentowski_2010, Deason_2011}. This is just one of many contradictory features of the DoS that make it highly problematic for $\Lambda$CDM \citep{Pawlowski_2021_Nature_Astronomy}. Other proposed formation scenarios in this framework \citep[like the accretion of galaxies along dark matter filaments;][]{Lovell_2011, Libeskind_2011} are statistically highly unlikely \citep[see, e.g.,][]{Pawlowksi_2014}. These scenarios are in any case already included in self-consistent cosmological simulations. In particular, EAGLE has been used to show that ``having most satellites accreted as a single group or along a single filament is unlikely to explain the MW DoS'' \citep{Shao_2018}.

\subsubsection{Alignment of the LMC with the DoS}
\label{subsubsec:Alignment of the LMC with the DoS}

Eight out of the 11 classical satellites co-orbit within the DoS and their orbital planes align within $20^{\circ}$ (Sculptor is counter-rotating). The orbital pole of the LMC (SMC) and the DoS normal have an angular distance of $\theta_{\mathrm{LMC}} = 19^{\circ}.2^{+0.4}_{-0.4}$ ($\theta_{\mathrm{SMC}} = 36^{\circ}.1^{+1.2}_{-1.1}$) based on combining proper motion measurements from the \emph{HST} and \emph{Gaia} \citep[see table~2 of][]{Pawlowski_2020}. In addition, the LMC and SMC co-orbit in the DoS along with most of the other classical satellites \citep[][]{Pawlowski_2012a}.

As demonstrated in the previous section, the DoS is unrelated to the formation of the MCs and MS in a first infall scenario. Thus, we quantify the likelihood of the LMC alignment with the DoS by assuming that the LMC falls towards the MW from a random direction independently of the DoS. We approximate that the orbital pole of the LMC aligns with that of the DoS to within $\theta \approx 20^{\circ}$. The likelihood that the orbital pole of an infalling satellite from a random direction aligns with the DoS within an angular distance $\theta$ and orbits in the same direction as most of the other satellites is given by
\begin{eqnarray}
    P \left( <\theta \right) ~=~ \frac{1 - \cos \theta}{2} \, ,
    \label{eq:likelihood_DoS_alignment}
\end{eqnarray}
which is about $0.030$ ($2.17 \sigma$) for $\theta = 20^{\circ}$ \citep[see also section~4.3 of][]{Pawlowski_2012a}. 

To calculate the combined likelihood of this alignment and the high phase-space density of the MCs, we add up the corresponding $\chi$ values in quadrature. 
\begin{eqnarray}
    \chi_{\mathrm{tot}}^{2} ~&=&~ \chi_{\mathrm{MCs}}^{2} + \chi_{\mathrm{LMC-DoS}}^{2} \, .
    \label{eq:combined_chisq}
\end{eqnarray}
The likelihood of a higher $\chi_{\mathrm{tot}}^{2}$ for two degrees of freedom is
\begin{eqnarray}
    P ~=~ \exp \left( -\frac{\chi_{\mathrm{tot}}^2}{2} \right).
    \label{eq:combined_Pvalue}
\end{eqnarray}
This is $P = 3.90\times 10^{-5}$ for $\chi_{\mathrm{tot}}^{2} = 3.95^2 + 2.17^2 = 20.30$ for the TNG50-1 simulation. Converting this $p$-value to an equivalent number of standard deviations for a single Gaussian variable by using Equation~\ref{P_chi} results in a $4.11\sigma$ tension. Thus, the MCs alone are roughly as problematic for the $\Lambda$CDM framework as any individual DoS in the LG, as summarized in Table~\ref{tab:list_problems_for_LCDM}. The results for the other simulations are summarized in Table~\ref{tab:results_LMC_SMC_analogues}.

\begin{table}
  \centering
    %\resizebox{1.1\linewidth}{!}{
    \begin{tabular}{lll}
    \hline
    Problem for $\Lambda$CDM & Frequency & Significance \\ \hline
    MW satellite plane & $~~~3.92 \times 10^{-4}$ & $~~~3.55\sigma$ \\
    M31 satellite plane & $~~~3.87 \times 10^{-4}$ & $~~~3.55\sigma$ \\
    NGC 3109 backsplash & $<7.56 \times 10^{-5}$ & $>3.96\sigma$ \\
    Phase-space density of MCs & $~~~7.81 \times 10^{-5}$ & $~~~3.95\sigma$ \\
    Phase-space density of MCs & \multirow{2}{*}{$~~~3.90 \times 10^{-5}$} & \multirow{2}{*}{$~~~4.11\sigma$} \\
     + MW DoS & & \\ \hline
    \end{tabular}%}
    \caption{Level of tension of different LG observations with the $\Lambda$CDM framework. The frequency of the MW DoS in $\Lambda$CDM is taken from section~4.2 of \citet{Pawlowski_2020}, who considered 2548 MW-like galaxies and found one with a similar satellite system. We therefore assume that its likelihood in $\Lambda$CDM is $1/2548 = 3.92 \times 10^{-4}$. For M31, we use figure~2 of \citet{Ibata_2014}, which shows that 3 out of 7757 M31-like galaxies in the MS-II have a satellite system similar to that of M31. The likelihood of the high mass and distance of the backsplash galaxy NGC~3109 in $\Lambda$CDM has been quantified in \citet{Banik_2021}, who found no such analogues in 13225 galaxies similar to the MW or M31 in the TNG300-1 simulation in terms of mass and distance \citep[see also][]{Osipova_2023}. This implies a likelihood below $1/13225 = 7.56 \times 10^{-5}$. The last two rows give the significance of the MCs as quantified using TNG50-1 in Sections~\ref{subsubsec:Phase-space density} and \ref{subsec:MCs and VPoS}.}
  \label{tab:list_problems_for_LCDM}
\end{table}

The flattened distribution and the phase-space correlation of the LG satellite galaxies strongly suggest that they are TDGs. Galaxies formed from tidal debris \citep[e.g.,][]{Mirabel_1992} have to lack dark matter due to their shallow gravitational potential and the high velocity dispersion of the CDM component \citep{Barnes_Hernquist_1992, Wetzstein_2007, Ploeckinger_2018, Haslbauer_2019}. However, the dwarf galaxies of the LG have high internal velocity dispersions, implying stronger self-gravity than what can be obtained from the baryonic matter alone \citep{McGaugh_2010, McGaugh_2013a, McGaugh_2013b} if Newtonian gravity is applied \citep[e.g.,][]{Kroupa_2015}.

\subsection{Total mass of LMC analogues}
\label{subsubsec:Total mass of LMC analogues}

\begin{figure*}
    \includegraphics[width=\columnwidth]{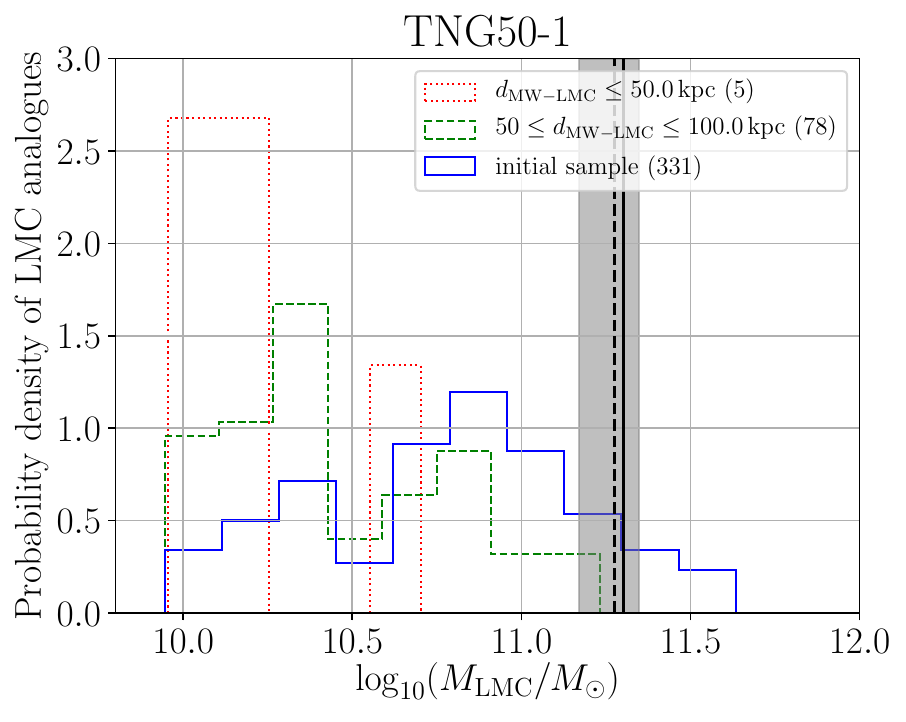}
    \includegraphics[width=\columnwidth]{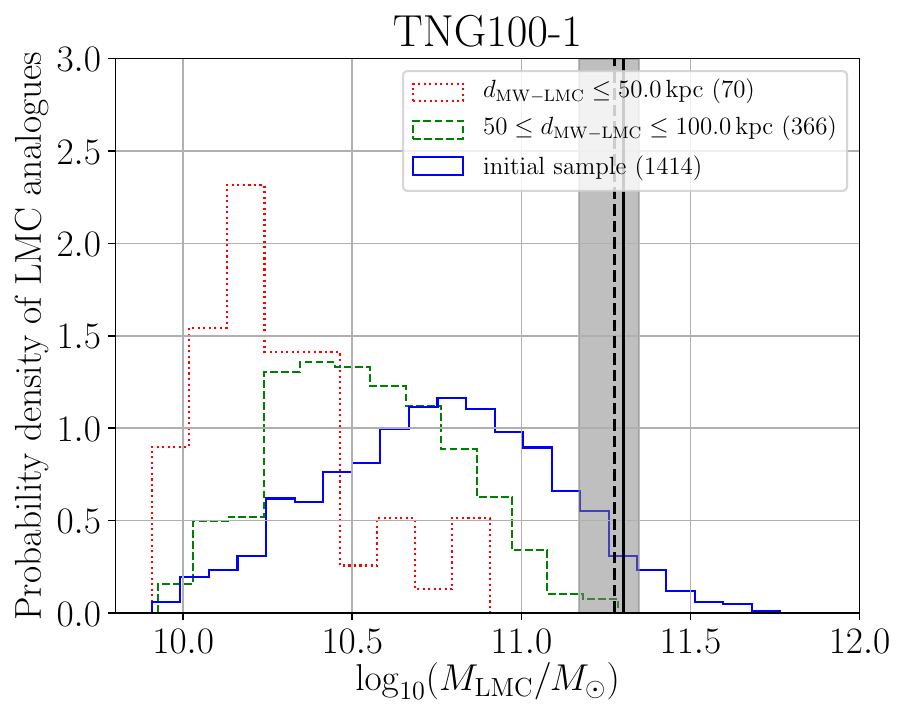}
    \includegraphics[width=\columnwidth]{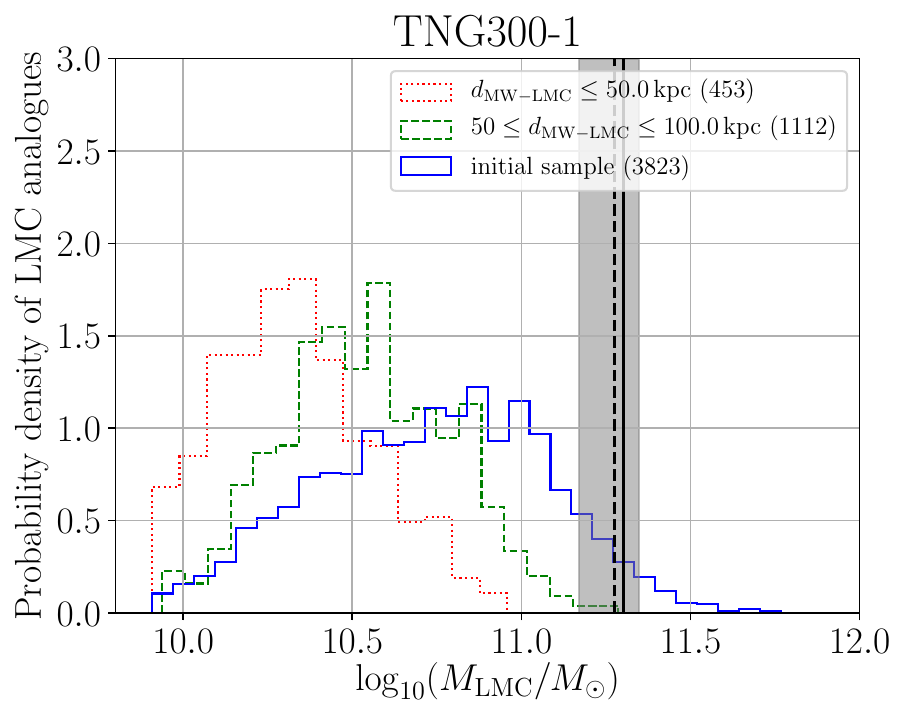}
    \includegraphics[width=\columnwidth]{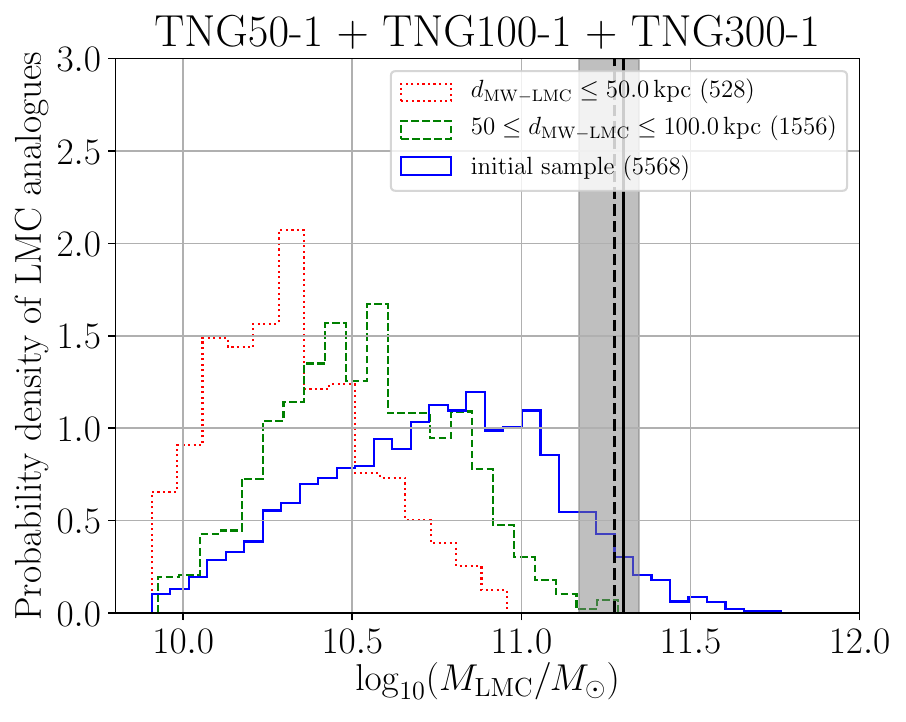}
    \caption{Histograms showing the total mass of LMC analogues with different distance criteria in TNG50-1 (\emph{top left}), TNG100-1 (\emph{top right}), and TNG300-1 (\emph{bottom left}), with the combined sample shown in the \emph{bottom right} panel. The solid blue histograms refer to the initial sample of analogues to the LMC (Section~\ref{subsec:Selection criteria of MCs-analogues}). The dotted red and dashed green histograms show the results when we also require the LMC to be within 50~kpc or between $50-100$~kpc, respectively, of the MW analogue. The sample sizes of the different distributions are given in brackets in the panel legends. The dashed vertical line and grey shaded region mark the total mass of the LMC and its uncertainty as deduced from five Galactic stellar streams near the LMC, which give $M_{\mathrm{LMC}}^{\mathrm{dyn}} = 18.8^{+3.5}_{-4.0} \times 10^{10}\,M_{\odot}$ \citep{Shipp_2021}. The solid vertical line shows the total mass of the LMC expected from abundance matching ($M_{\mathrm{LMC}} \approx 2 \times 10^{11} \, M_{\odot}$). Constraining the total mass of the LMC with the LG timing argument gives similar results (see the text). Table~\ref{tab:Total_mass_LMC} lists the number of LMC analogues with a total mass above the $1\sigma$ lower limit of the analysis by \citet{Shipp_2021}.}
    \label{figure_results_LMC_masses}
\end{figure*}

Throughout the analysis of Section~\ref{sec:Results}, we have imposed a lower limit on the stellar masses of the MCs but have not considered the total masses of analogues to them. Since the LMC is much more massive than the SMC, we focus on the LMC in the following analysis. \citet{Shipp_2021} measured a total LMC mass of $M_{\mathrm{LMC}}^{\mathrm{dyn}} = 18.8^{+3.5}_{-4.0} \times 10^{10}\,M_{\odot}$ by using five southern hemisphere stellar streams as direct dynamical tracers of the underlying gravitational potential. Their result is similar to other studies of the LMC and its surroundings \citep[see figure~17 of][]{Koposov_2023}, so we compare it to the distribution of the total LMC mass ($M_{\mathrm{LMC}}$)\footnote{$M_{\mathrm{LMC}}$ here refers to the total mass of all particles and cells which are bound to this subhalo as listed in the \textsc{subfind} subhalos catalogue (\url{https://www.tng-project.org/data/docs/specifications/\#sec2b} [23.09.2024]).} in Figure~\ref{figure_results_LMC_masses}, which has four panels showing results from TNG50-1, TNG100-1, TNG300-1, and the combination of all three. In each panel, results are shown for three samples depending on the imposed distance criteria. The initial LMC sample of Section~\ref{subsec:Selection criteria of MCs-analogues} shows a peak at $\log_{10} \left( M_{\mathrm{LMC}}/M_{\odot} \right) \approx 10.8-10.9$ in all three runs, with the distribution extending to $\approx 11.8$ in the TNG100-1 and TNG300-1 runs (see the solid blue histograms). This is consistent with the halo mass of the LMC expected from abundance matching \citep[$\log_{10} \left( M_{\mathrm{LMC}}/M_{\odot} \right) \approx 11.3$; see e.g.][]{BoylanKolchin_2010, Moster_2013, Dooley_2017, Shao_2018_LMC}.

Selecting only LMC analogues with $50 \, \mathrm{kpc} \le d_{\mathrm{MW-LMC}} \le 100 \, \mathrm{kpc}$ systematically shifts the $M_{\mathrm{LMC}}$ distribution to lower masses, with the peak now around $\log_{10} \left( M_{\mathrm{LMC}}/M_{\odot} \right) \approx 10.5$. Combining all three simulation runs gives that nine out of 1556 LMC analogues have a total mass equal to or larger than the $1\sigma$ lower limit of the tidal stream analysis conducted by \citet{Shipp_2021}. This frequency corresponds to a $2.76 \sigma$ tension.

\begin{table*}
  \centering
	\resizebox{\linewidth}{!}{\begin{tabular}{lllll}
	\hline
    Sample & Simulation & Number of analogues & Frequency & Significance \\ \hline 
    Initial sample of analogues to the LMC (Section~\ref{subsec:Selection criteria of MCs-analogues}) 
    & TNG50-1: & $54$ & $54/331$ & $1.39 \sigma $ \\
    $+M_{\mathrm{LMC}} \ge (18.8 - 4.0) \times 10^{10} \, M_\odot$ & TNG100-1: & $163$ & $163/1414$ & $1.57 \sigma $ \\
    & TNG300-1: & $354$ & $354/3823$ & $1.68 \sigma $ \\
    & All runs: & $571$ & $571/5568$ & $1.63 \sigma $ \\ \hline
   Sample of analogues to the LMC with $50 \, \mathrm{kpc} \le d_{\mathrm{MW-LMC}} \le 100 \, \mathrm{kpc}$ & TNG50-1: & $2$ & $2/78$ & $2.23 \sigma $ \\
   $+M_{\mathrm{LMC}} \ge (18.8 - 4.0) \times 10^{10} \, M_\odot$  & TNG100-1: & $3$ & $3/366$ & $2.64 \sigma $  \\
     & TNG300-1: & $4$ & $4/1112$ & $2.91 \sigma $ \\
    & All runs: & $9$ & $9/1556$ & $2.76 \sigma $ \\ \hline 

   Sample of analogues to the LMC with $d_{\mathrm{MW-LMC}} \le 50 \, \mathrm{kpc}$ & TNG50-1: & $0$ & $<1/5$ & $>1.28 \sigma $\\
   $+M_{\mathrm{LMC}} \ge (18.8 - 4.0) \times 10^{10} \, M_\odot$ & TNG100-1: & $0$ & $<1/70$ & $>2.45 \sigma $ \\
   & TNG300-1: & $0$ & $<1/453$ & $>3.06 \sigma $\\
    & All runs: & $0$ & $<1/528$ & $>3.11 \sigma $ \\	\hline
    
    Initial sample of analogues to the LMC (Section~\ref{subsec:Selection criteria of MCs-analogues}) & All runs: & $0$ & $<1/5568$ & $>3.75 \sigma $\\
   $+M_{\mathrm{LMC}} \ge (18.8 - 4.0) \times 10^{10} \, M_\odot$ \& $d_{\mathrm{MW-LMC}} \le 50 \, \mathrm{kpc}$ & & & & \\	\hline
   
    Initial sample of analogues to the LMC (Section~\ref{subsec:Selection criteria of MCs-analogues}) & All runs: & $0$ & $<3.62 \times 10^{-7}$ & $>5.09 \sigma $\\
   $+M_{\mathrm{LMC}} \ge (18.8 - 4.0) \times 10^{10} \, M_\odot$ \& $d_{\mathrm{MW-LMC}} \le 50 \, \mathrm{kpc}$ & & & & \\
   +Extrapolation of $f_{\mathrm{MCs}}$ distribution using quadratic fit (TNG50-1)& & & & \\ \hline
   	\end{tabular}}
	\caption{Statistics of analogues to the LMC (i.e., systems with a satellite analogous to the LMC) in the redshift range $0 \le z \le 0.26$ in the TNG50-1, TNG100-1, and TNG300-1 simulation with a total mass higher than the $1\sigma$ lower limit of the observationally inferred LMC mass \citep[$M_{\mathrm{LMC}}^{\mathrm{dyn}} = 18.8^{+3.5}_{-4.0} \times 10^{10}\,M_\odot$;][]{Shipp_2021}. We show results for different distance criteria. Similarly to Table~\ref{tab:results_LMC_SMC_analogues}, we list the number of analogues, their frequency relative to the total number of selected MW-like analogues, and the corresponding equivalent number of standard deviations for a single Gaussian variable (Equation~\ref{P_chi}). The distribution of the total LMC mass for each of these three samples is presented in Figure~\ref{figure_results_LMC_masses}. The last two parts of the table list the statistical analysis of Figure~\ref{figure_results_LMC_distances_masses}.}
  \label{tab:Total_mass_LMC}
\end{table*}

Imposing that $d_{\mathrm{MW-LMC}} \le 50.0 \, \mathrm{kpc}$ shifts the distribution to even lower $M_{\mathrm{LMC}}$. The maximum mass of these LMC analogues is $\log_{10} \left( M_{\mathrm{LMC}}/M_{\odot} \right) \approx 10.96$ (obtained in TNG300-1). This is still lower than the $1\sigma$ lower limit of the tidal stream analysis of \citet{Shipp_2021}, implying a null detection of sufficiently massive LMC analogues. The results for the different selection criteria and simulation runs are summarized in Table~\ref{tab:Total_mass_LMC}.

\begin{figure}
    \includegraphics[width=\columnwidth]{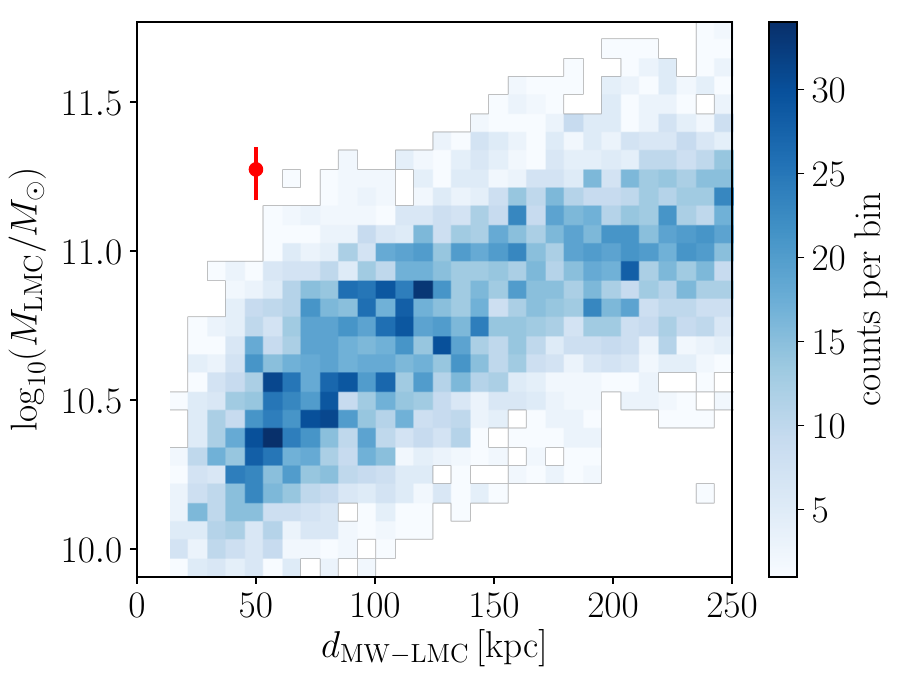}
    \caption{Distribution of the MW-LMC distance and the total LMC mass of analogues in the combined sample of the three TNG runs (5568 objects). The red dot with vertical error bar refers to the observed $d_{\mathrm{MW-LMC}} = 50.0 \,\mathrm{kpc}$ and the total LMC mass deduced from five Galactic stellar streams near the LMC \citep[these give $M_{\mathrm{LMC}} = 18.8^{+3.5}_{-4.0} \times 10^{10}\,M_{\odot}$;][]{Shipp_2021}. None of the 5568 simulated objects have a lower Galactocentric distance and higher mass than the LMC.}
    \label{figure_results_LMC_distances_masses}
\end{figure}

Our results show that LMC analogues within $50$~kpc\ but also those with Galactocentric distances of $50-100$ kpc are systematically less massive compared to the initial sample, where LMC analogues could be up to 250~kpc from their MW-like host galaxy (Section~\ref{subsec:Selection criteria of MCs-analogues}). This distance dependence is also evident in Figure~\ref{figure_results_LMC_distances_masses}, which shows the joint distribution of the MW-LMC distance and the total LMC mass. As expected from Figure~\ref{figure_results_LMC_masses}, the typical LMC mass systematically increases with its distance from the host galaxy. 

The null detection of such systems is probably due to dynamical friction between CDM halos: massive satellites rapidly merge with the host galaxy, making the MW-LMC configuration less likely for a more massive LMC (Section~\ref{subsec:Effect of dynamical friction}). As a result, none of the 5568 LMC analogues in our initial sample have a lower Galactocentric distance and higher total mass than estimated for the real LMC. This represents a tension of $>3.75\sigma$, which would no doubt rise further if we simply showed the mass-distance distribution of the most massive satellite around all MW analogues without \emph{a priori} requiring the LMC analogue to have a sufficiently high stellar mass. Since this estimate does not take into account the SMC, we can combine the tension evident in Figure~\ref{figure_results_LMC_distances_masses} with our previously reported $3.95\sigma$ tension from the high phase-space density of the MCs (Section~\ref{subsubsec:Phase-space density}), which results in a combined highly significant $>5.09\sigma$ tension for two degrees of freedom ($\chi_{\mathrm{tot}}^2 > 3.75^2 + 3.95^2$; Equation~\ref{eq:combined_Pvalue}).

\section{Conclusions}
\label{sec:Conclusion}

The MCs are the most massive and some of the closest satellites of the MW and have been known since ancient times, making them an interesting local Universe laboratory for cosmological and gravitational theories. Our proximity to the MCs allows us to constrain their orbits to high precision \citep[e.g.,][]{Kallivayalil_2013, GaiaCollaboration_2016, Gaia_2018a, Gaia_2018b}. In this contribution, we quantified the likelihood of a system resembling the MW and the MCs in the standard model of cosmology ($\Lambda$CDM) as simulated in the IllustrisTNG project \citep{Pillepich_2018, Nelson_2019}. We identified $1547$, $7360$, and $40075$ MW-like galaxies in the redshift range $0.0 \le z \le 0.26$ in TNG50-1, TNG100-1, and TNG300-1, respectively (Section~\ref{subsec:Selection criteria}). We then selected satellites within $250\,\mathrm{kpc}$ and ranked them according to their stellar mass. After that, we considered the most massive satellite of a halo as a possible LMC analogue if it has $M_{\star} \ge 1.5 \times 10^{9}\,M_\odot$ and $M_{\mathrm{total}}/M_{\star} > 5$. For the less massive SMC, we required that the second-most massive satellite (in terms of stellar mass) has $M_{\star} \ge 4.6 \times 10^{8}\,M_\odot$ and $M_{\mathrm{total}}/M_{\star} > 5$ (Section~\ref{subsec:Selection criteria of MCs-analogues}). In total, $147$, $454$, and $601$ MW-like galaxies have analogues to both MCs in this sense, corresponding to a frequency of $9.5\%$, $6.2\%$, and $1.5\%$ in TNG50-1, TNG100-1, and TNG300-1, respectively.

Using these galaxy samples, we investigated the formation and evolution of analogues to the MCs by applying different observational constraints. The unusual aspect of the MCs is that it is a closely interacting galactic system with a mutual separation of $24.5$~kpc and a low relative velocity of $90.8\,\mathrm{km\,s^{-1}}$, which implies a high specific phase-space density of $f_{\mathrm{MCs,obs}} = 9.10 \times 10^{-11} \,\mathrm{km^{-3}\,s^{3}\,kpc^{-3}}$. In particular, 3 out of 1202 systems have $f_{\mathrm{MCs}} \ge f_{\mathrm{MCs,obs}}$. However, none of the identified MCs analogues with $50~\mathrm{kpc} \le d_{\mathrm{MW-LMC}} \le 100~\mathrm{kpc}$ in the TNG runs have $f_{\mathrm{MCs}} \ge f_{\mathrm{MCs,obs}}$. This null detection implies a frequency of $<1/46$, $< 1/118$, and $<1/191$, which corresponds to a $> 2.29 \sigma$, $> 2.63 \sigma$, and $> 2.79 \sigma$ tension in the TNG50-1, TNG100-1, and TNG300-1 run, respectively. This analysis sets the lower limit to the tension by relying on the null detection of MCs analogues. Consequently, we performed an extrapolation of the cumulative $f_{\mathrm{MCs}}$ distribution in order to estimate the actual significance (Section~\ref{subsubsec:Phase-space density}). Extrapolating the simulated $f_{\mathrm{MCs}}$ distribution up to $f_{\mathrm{MCs,obs}}$ yields a tension of $3.95 \sigma$ (TNG50-1), $3.11\sigma$ (TNG100-1), and $2.82\sigma$ (TNG300-1). We emphasize that this extrapolation is a usual method applied in extreme value analysis to assess the likelihoods of outcomes when the numerical experiments are too costly to provide sufficient samples to cover the extreme values of interest. The discrepancy between the simulated and observed phase-space density of the MCs is likely not caused by the resolution limitations of the simulations (Appendix~\ref{subsubsec:Effect of resolution}). There are fewer MW-MCs systems in the lower resolution runs of the TNG50 and TNG100 simulations, which could be due to artificial disruption of interacting subhalos. The statistical significance of the tension between the observed and simulated specific phase-space density distribution is higher in the high-resolution run TNG50-1. Thus, a hierarchical clustering of two massive satellites in a narrow phase-space volume is an unlikely configuration within the $\Lambda$CDM framework, regardless of precisely what distance range is allowed for the LMC. Testing cosmological theory using the phase-space density of the MW-LMC-SMC system is applied here for the first time. Notably, this configuration is very comparable to the similar situation observed in the nearby M81 group of galaxies \citep{Oehm_2017}, one of the closest galaxy groups to the LG.

By tracing the three systems with $f_{\mathrm{MCs}} \ge f_{\mathrm{MCs,obs}}$ in our initial MCs sample back through cosmic time, we found that they are accreted together towards the MW-like galaxy at late times (Section~\ref{subsubsec:Individual analogues}). This so-called `first infall scenario' is thus the most likely formation process of the MCs in a $\Lambda$CDM universe, in agreement with previous studies \citep[e.g.,][]{BoylanKolchin_2011}. The low LMC-SMC relative velocity is unusual for a scenario where both MCs fell in from large distances and only encountered each other in the last few Gyr. This is one reason for the high observed phase-space density of the MCs compared to their analogues found in $\Lambda$CDM simulations.

The orbital pole of the LMC (SMC) and the DoS normal have an angular distance of $19^\circ.2^{+0.4}_{-0.4}$ \citep[$36^\circ.1^{+1.2}_{-1.1}$;][]{Pawlowski_2020}. As demonstrated in Section~\ref{subsubsec:Satellites of the LMC}, the LMC cannot bring in enough satellites to populate the satellite plane, a finding that is consistent with previous studies \citep[e.g.,][]{SantosSantos_2021}. Thus, the DoS is physically unrelated to the proposed recent first infall of the MCs in the $\Lambda$CDM framework. We estimate that the alignment of the LMC and DoS has a likelihood of $0.030$ ($2.17 \sigma$) if the LMC falls towards the MW from a random direction on the sky. Adding this $\chi^{2}$ value to that of the phase-space density of the MCs increases the tension to $4.11 \sigma$ for two degrees of freedom in TNG50-1, assuming the DoS \emph{a priori}. However, it was previously shown that the existence of the DoS around the MW causes a $3.55\sigma$ tension based on TNG100-1 \citep{Pawlowski_2020}. Consequently, although most physical properties of the MCs and the MS can be explained in non-cosmological simulations \citep{Lucchini_2020, Lucchini_2021}, their formation and alignment with the DoS remain a challenge in cosmological $\Lambda$CDM simulations. Looking slightly further afield, the fact that the satellite populations of the three nearest major host galaxies (the MW, M31, and Centaurus A) all reveal a DoS is in $5.27\sigma$ tension with the $\Lambda$CDM model \citep{Asencio_2022}, even without considering that the proper motions of several M31 satellites align with its dominant satellite plane \citep{Sohn_2020, Casetti_2024}. Similarly detailed studies are currently not possible around more distant hosts, though some progress in this regard has recently been achieved around NGC~4490 \citep{Pawlowski_2024} and a few other hosts \citep[as reviewed in][]{Pawlowski_2021}.

Another problematic aspect of the LMC is the fact that it is only 50~kpc away but has a Newtonian dynamical mass of $M_{\mathrm{LMC}} = 18.8^{+3.5}_{-4.0} \times 10^{10}\,M_{\odot}$ according to an analysis of five tidal streams near it \citep{Shipp_2021}. Timing argument analyses of the LG give similar results (see Section~\ref{subsubsec:Total mass of LMC analogues}). Figure~\ref{figure_results_LMC_distances_masses} demonstrates that the LMC is quite massive given its distance of only $50$~kpc from the MW. Indeed, none of the 5568 LMC analogues in terms of stellar mass have a lower Galactocentric distance and higher total mass than the above estimates for the actual LMC, even if we adopt the $1\sigma$ lower limit on its mass. The sample size would rise further if we did not impose the condition on $M_{\star}$ a priori, indicating a tension of $>3.75\sigma$. We attribute the lack of sufficiently massive and nearby satellites to dynamical friction \citep{Kroupa_2015, Oehm_2024}. Combining this with the $3.95\sigma$ tension from the phase-space density of the MCs results in a $>5.09\sigma$ tension for two degrees of freedom.

The tensions caused by the phase-space density of the MCs and the alignment of the LMC with the DoS can be alleviated if dynamical friction on their CDM halos is absent (Section~\ref{subsec:Effect of dynamical friction}). This evidence for a lack of dynamical friction on galactic scales is consistent with previous studies \citep[e.g.,][]{Angus_2011, Kroupa_2015, Oehm_2017, Roshan_2021_disc_stability, Roshan_2021_bar_speed, Haslbauer_2022, Oehm_2024}. Altering the detailed properties of the dark matter particle might help with this issue. Some models along these lines are superfluid dark matter \citep[SFDM;][]{Berezhiani_2015, Berezhiani_2018, Berezhiani_2019} and self-interacting dark matter \citep[SIDM;][]{Spergel_2000, Ren_2019}, which can reduce the dark matter density in the central few kpc of a galaxy and thus the resulting dynamical friction on those scales. While that might be particularly helpful at avoiding excessive slowdown of bar pattern speeds, any Newtonian model of the Galaxy must have a dark matter density profile set by its rotation curve. This means the dark matter density profile on the 100~kpc scale relevant to the MCs would be much the same in any Newtonian gravity plus dark matter interpretation of the observed rotation curve anomalies in galaxies. However, dynamical friction arises on the much smaller scale of the wake in the dark matter generated by the passage of a satellite. As a result, it is possible that dynamical friction would be reduced in these scenarios, as indeed seems to be the case with fuzzy dark matter \citep[FDM;][]{Lancaster_2020, Buehler_2023, Vitsos_2023}. Cosmological simulations in these scenarios would need to be analysed similarly to this work to assess if they can better match the observed properties of the MCs, in particular their small separation and relative velocity (Equation~\ref{eq:specific_phase_space_density_obs}). However, the fact that FDM particles would need to account for the large dark matter content of the faint satellite galaxies with spatial scales $\ll 1$~kpc \citep[including field galaxies where tides cannot explain the elevated velocity dispersions; see figure~16 of][and references therein]{Asencio_2022} means that FDM particles would behave as ordinary CDM on scales relevant to the orbits of the MCs.

Dynamical friction on galactic scales would be greatly reduced in Milgromian dynamics \citep[MOND;][]{Milgrom_1983} due to the lack of CDM halos \citep[for extensive reviews of MOND, see][]{Famaey_2012, Banik_2022_MONDreview}. The absence of DM halos around dwarf galaxies has recently been reported in the Fornax galaxy cluster based on Newtonian tidal stability arguments \citep{Asencio_2022}, creating significant tension with the $\Lambda$CDM model (though MOND is in good agreement with the observations). In MOND, the fact that the Galactocentric velocities of the MCs are well below the Galactic escape velocity at the relevant distance \citep{Banik_2018_escape} implies that the MCs can either be primordial galaxies but not on a first infall, or they could be TDGs formed by a past MW-M31 flyby \citep{Zhao_2013}. The flyby scenario would also provide a natural explanation for the DoSs around the MW and M31 \citep{Banik_2018, Banik_2022_SP, Bilek_2018}, both of which individually cause a $3.55\sigma$ tension with the $\Lambda$CDM framework \citep{Ibata_2014, Pawlowski_2020}. Self-consistent hydrodynamical cosmological MOND simulations are needed to rigorously quantify the likelihood of the observed LG configuration in MOND, as presented in this contribution for the $\Lambda$CDM framework.

The here-performed extensive data mining of the highest-resolution cosmological hydrodynamical structure formation simulations leads to the result that they cannot form MW-LMC-SMC triple systems analogous to the real system. It can be argued that one merely needs to do more supercomputer simulations to finally find a triple system similar to the observed one. However, \citet{Oehm_2024} conduct an extensive study searching for orbital solutions by integrating the presently observed configuration backwards in time, explicitly taking into account the Chandrasekhar dynamical friction that would arise in $\Lambda$CDM. Their results demonstrate that solutions do not exist because the orbital decay due to dynamical friction is too significant given the CDM halos predicted by $\Lambda$CDM structure formation theory. In particular, it becomes impossible to have a prior close interaction between the MCs $1-4$~Gyr ago, which is necessary to form the MS \citep[see our Section~\ref{subsubsec:Individual analogues} and the non-cosmological simulations of][]{Lucchini_2020, Lucchini_2021}. The results obtained here are thus consistent with the results from the orbit calculations of \citet{Oehm_2024}, implying that the observed MW-LMC-SMC triple system cannot exist in dark matter models. Assuming that the main difference in MOND is the lack of dynamical friction on the MCs as they orbit each other and the MW, the work of \citet{Oehm_2024} suggests that orbital solutions can be readily obtained in MOND.

As a final remark, we note that large galaxy surveys should be complemented by tests of cosmological models and gravitational theories in the local Universe, the only region where we have access to many types of precise yet model-independent data thanks to missions like \emph{Gaia} \citep[cf.][]{Kroupa_2022}.

\section*{Funding}

IB is supported by Royal Society University Research Fellowship 211046 and was supported by Science and Technology Facilities Council grant ST/V000861/1, which also partially supports HZ. IB acknowledges support from a ``Pathways to Research'' fellowship from the University of Bonn. PK thanks the Deutscher Akademischer Austauschdienst-Eastern European Exchange Programme at the University of Bonn for support. EA acknowledges
support through a teaching assistantship by the Helmholtz-Institut f\"ur Strahlen- und Kernphysik (HISKP). The IllustrisTNG simulations were undertaken with computational time awarded by the Gauss Centre for Supercomputing (GCS) under GCS Large-Scale Projects GCS-ILLU and GCS-DWAR on the GCS share of the supercomputer Hazel Hen at the High Performance Computing Center Stuttgart (HLRS), as well as on the machines of the Max Planck Computing and Data Facility (MPCDF) in Garching, Germany.

\begin{acknowledgments} %Based on https://journals.aas.org/aastexguide/#acknowledgments

IB thanks Harry Desmond for advice about comparing models with different numbers of free parameters. The authors thank the anonymous referees for their helpful comments.

\end{acknowledgments}

\section*{Data availability}

No new data were created or analysed in this study.

\section*{Conflicts of interest}

The authors declare no conflicts of interest.

\section*{Author contributions}

Formal analysis: Moritz Haslbauer; Investigation, Moritz Haslbauer and Indranil Banik; Methodology, Moritz Haslbauer, Indranil Banik, Pavel Kroupa, Hongsheng Zhao and Elena Asencio; Supervision, Indranil Banik and Pavel Kroupa; Writing $-$ original draft, Moritz Haslbauer; Writing $-$ review \& editing, Indranil Banik, Pavel Kroupa, Hongsheng Zhao and Elena Asencio.

% Manual protocol for references (use only as a last resort).
% This method is tedious and prone to error if you have lots of references.
%\begin{thebibliography}{99}
%\end{thebibliography}

%%%%%%%%%%%%%%%%% APPENDICES %%%%%%%%%%%%%%%%%%%%%

%Example appendix from IB that actually works.
%Put \end{document} command at the end.

\begin{appendix}
\section{Effect of resolution}
\label{subsubsec:Effect of resolution}

\begin{figure*}
    \includegraphics[width=5.8cm]{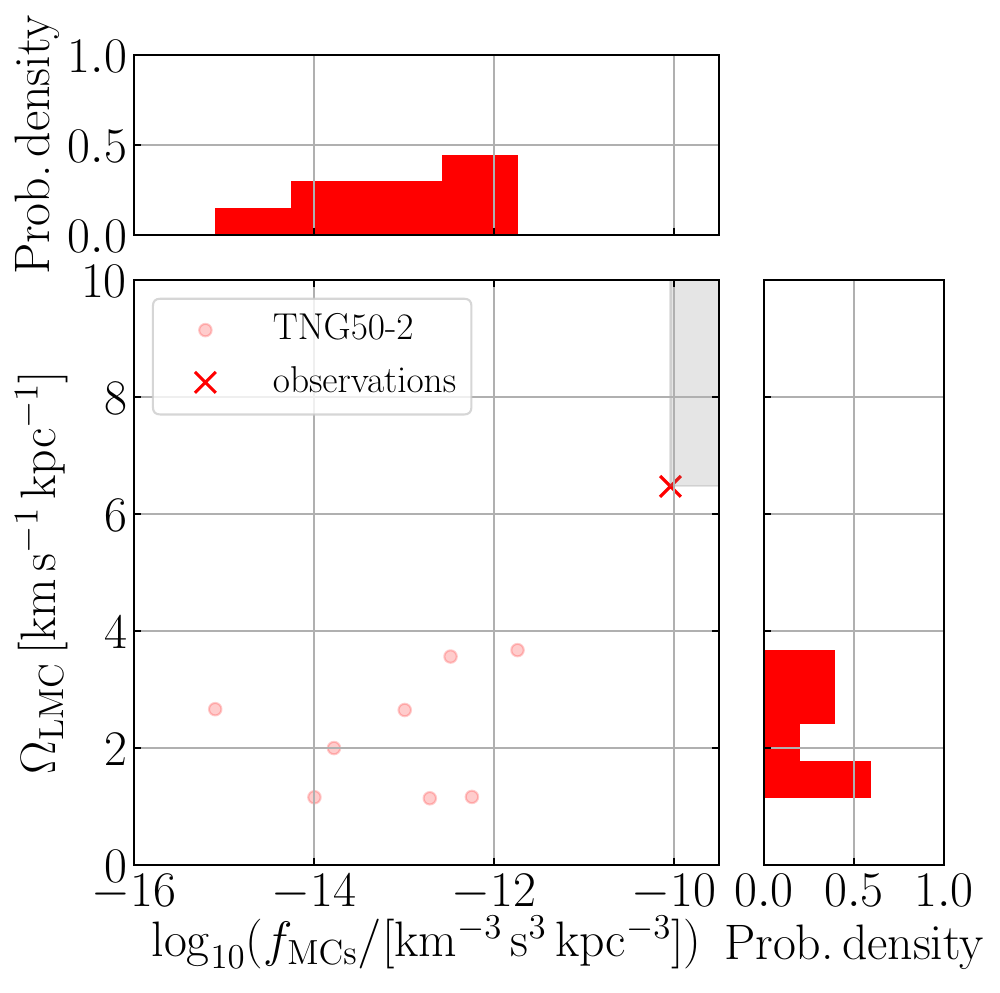}
    \includegraphics[width=5.8cm]{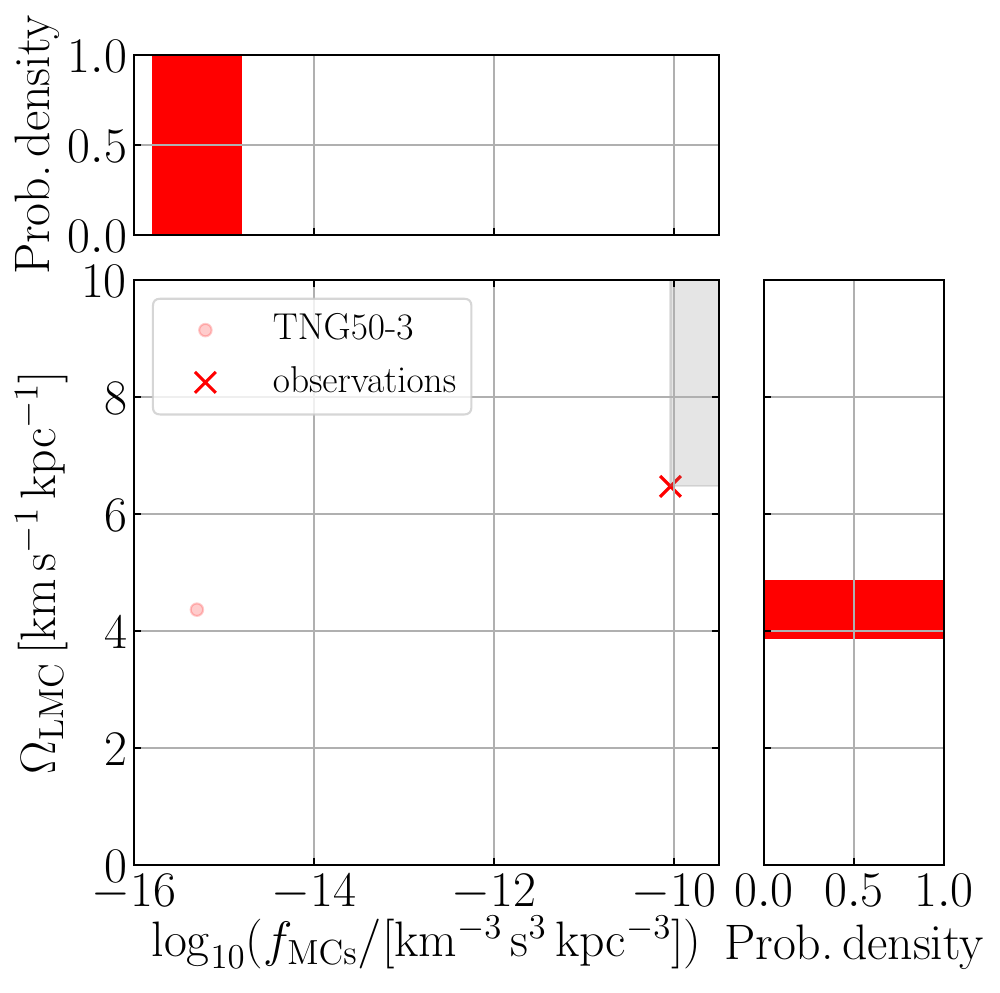}
    \includegraphics[width=5.8cm]{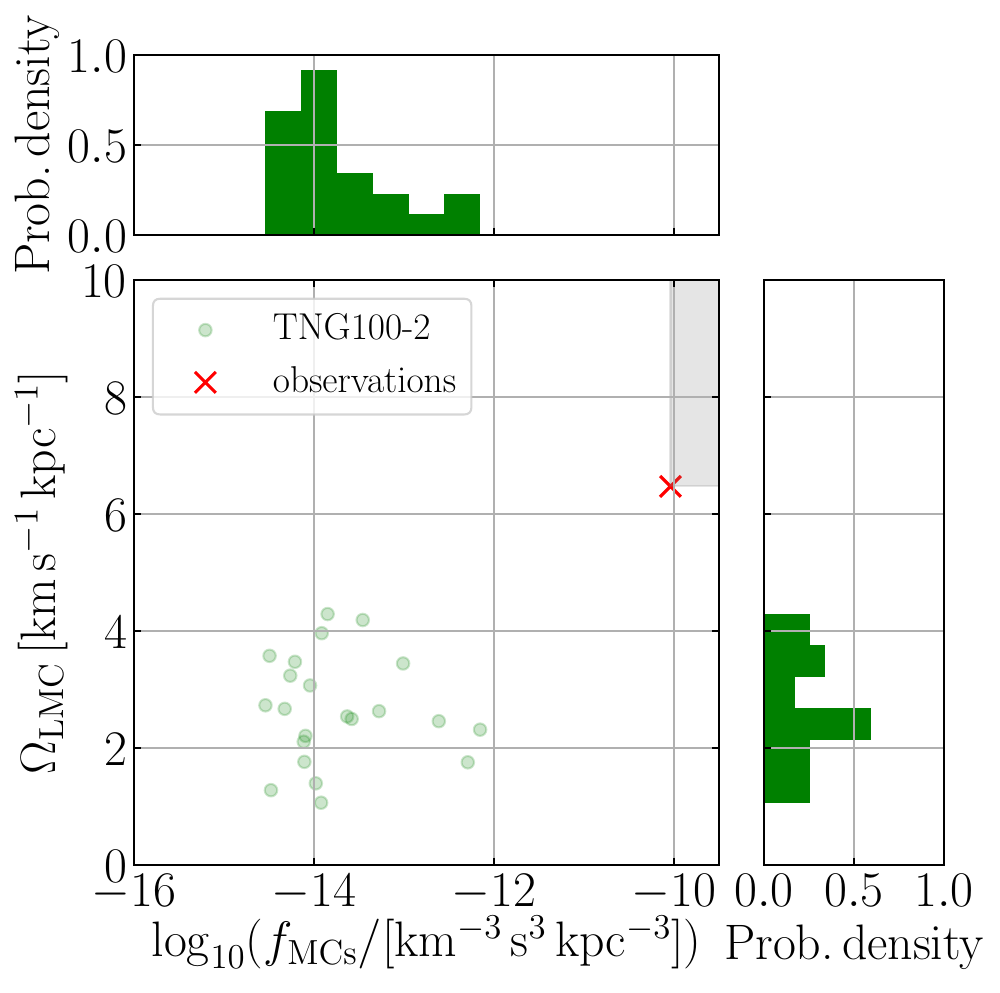}
    \caption{Same as Figure~\ref{figure_phase_space}, but for the low-resolution runs TNG50-2 (8 objects; \emph{left}), TNG50-3 (1 object; \emph{middle}), and TNG100-2 (22 objects; \emph{right}).}
    \label{figure_phase_space_low_resolution}
\end{figure*}

\begin{figure*}
    \includegraphics[width=9cm]{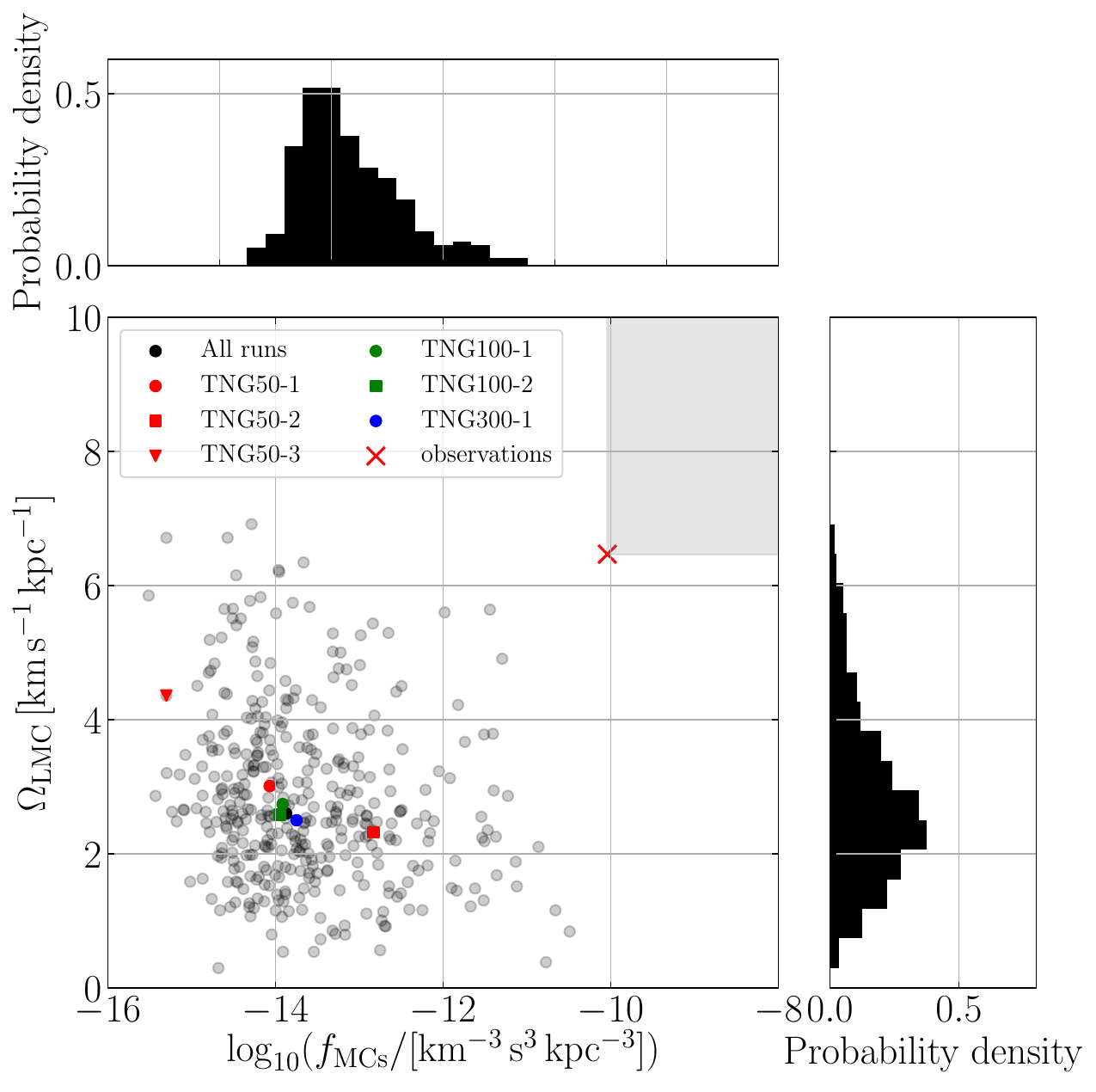}
    \includegraphics[width=9cm]{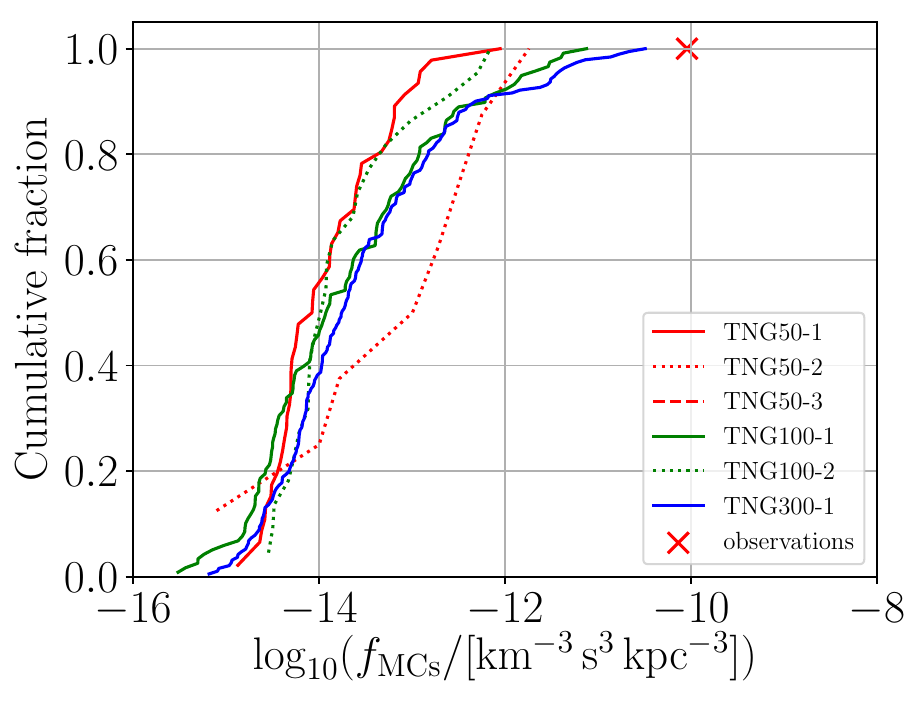}
    \caption{\emph{Left panel}: Similar to Figure~\ref{figure_phase_space}, but here the black dots show the results of TNG50-1 (46 objects), TNG50-2 (8 objects), TNG50-3 (one object), TNG100-1 (118 objects), TNG100-2 (22 objects), and TNG300-1 (191 objects). The coloured symbols mark the medians of the different resolution runs: $\log_{10} \left( f_{\mathrm{MCs}}/\mathrm{\left[ km^{-3}\,s^{3}\,kpc^{-3} \right]} \right) = -14.07$ (TNG50-1), $-12.83$ (TNG50-2), $-15.30$ (TNG50-3), $-13.92$ (TNG100-1), $-13.95$ (TNG100-2), and $-13.75$ (TNG300-1). \emph{Right panel:} Cumulative distribution of the specific phase-space density of the MCs ($f_{\mathrm{MCs}}$; Equation~\ref{eq:specific_phase_space_density}). In both panels, the red cross shows the observed value of $f_{\mathrm{MCs,obs}} = 9.10 \times 10^{-11} \, \mathrm{km^{-3}\,s^{3}\,kpc^{-3}}$.}
    \label{figure_phase_space_resolution_test}
\end{figure*}

The distributions of $f_{\mathrm{MCs}}$ (Equation~\ref{eq:specific_phase_space_density}) and $\Omega_{\mathrm{LMC}}$ (Equation~\ref{eq:velocity_distance_LMC}) of MCs analogues with $50 \, \mathrm{kpc} \le d_{\mathrm{MW-LMC}} \le 100\,\mathrm{kpc}$ in the three low-resolution runs TNG50-2, TNG50-3, and TNG100-2 are shown in Figure~\ref{figure_phase_space_low_resolution}. The observed phase-space density of the MCs ($\log_{10} \left( f_{\mathrm{MCs,obs}}/\mathrm{\left[ km^{-3}\,s^{3}\,kpc^{-3} \right]} \right) \approx -10.04$) is significantly higher than expected from the TNG simulations, which have $f_{\mathrm{MCs}}$ distributions with medians of $\log_{10} \left( f_{\mathrm{MCs}}/\mathrm{\left[ km^{-3}\,s^{3}\,kpc^{-3} \right]} \right) = -14.07$ (TNG50-1), $-12.83$ (TNG50-2), $-15.30$ (TNG50-3; but this refers to only one data point), $-13.92$ (TNG100-1), $-13.95$ (TNG100-2), and $-13.75$ (TNG300-1), as visualized in the left panel of Figure~\ref{figure_phase_space_resolution_test}. Thus, TNG50-1, TNG100-1, TNG100-2, and TNG300-1 all have medians that cluster within the same region: $\log_{10} \left( f_{\mathrm{MCs}}/\mathrm{\left[ km^{-3}\,s^{3}\,kpc^{-3} \right]} \right) \approx -14$ and $\Omega_{\mathrm{LMC}} \approx 2.7\,\mathrm{km\,s^{-1}\,kpc^{-1}}$. The highest resolution run TNG50-1 has an even lower median than the lower resolution run TNG50-2, implying that increasing the resolution does not increase the phase-space density of MCs analogues in the simulations. This conclusion is supported by the right panel of Figure~\ref{figure_phase_space_resolution_test}, which shows that the cumulative $f_{\mathrm{MCs}}$ distribution is very similar between different TNG runs. Moreover, the cumulative distribution of TNG50-1 is shifted to even lower values than in the lower resolution runs TNG100-1 and TNG300-1, which increases therewith the tension in TNG50-1. Consequently, we conclude that the discrepancy between the observed and simulated phase-space configuration of the MCs is likely not caused by the numerical limitations of the assessed TNG50-1 run $-$ if anything, improving the resolution worsens the problem.

\section{Comparison of polynomial fits to the \texorpdfstring{$\lowercase{f}_{\mathrm{MC\lowercase{s}}}$}{f\_MCs} distribution}
\label{subsec:Comparing polynomial fits}

\begin{figure*}
    \includegraphics[width=8.7cm]{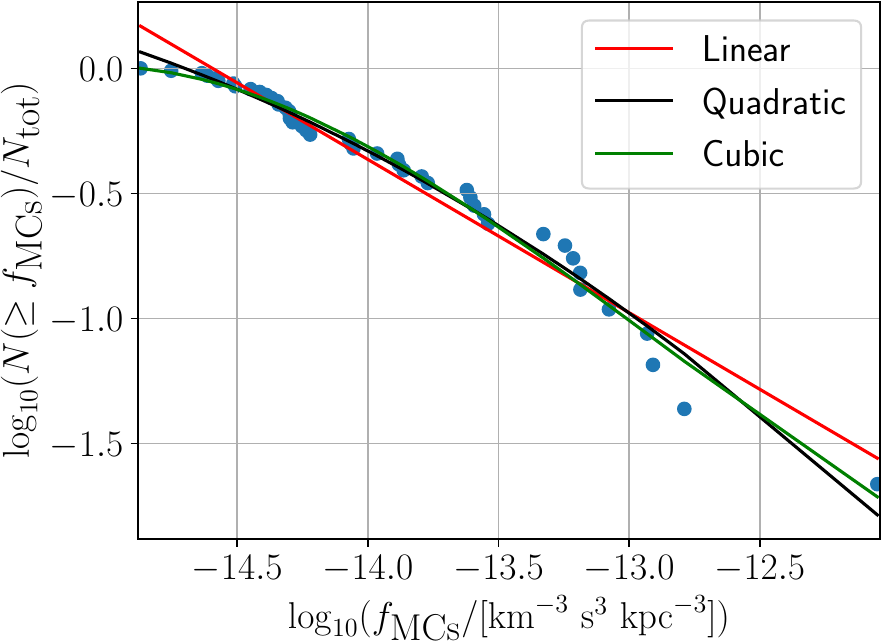}
    \includegraphics[width=8.7cm]{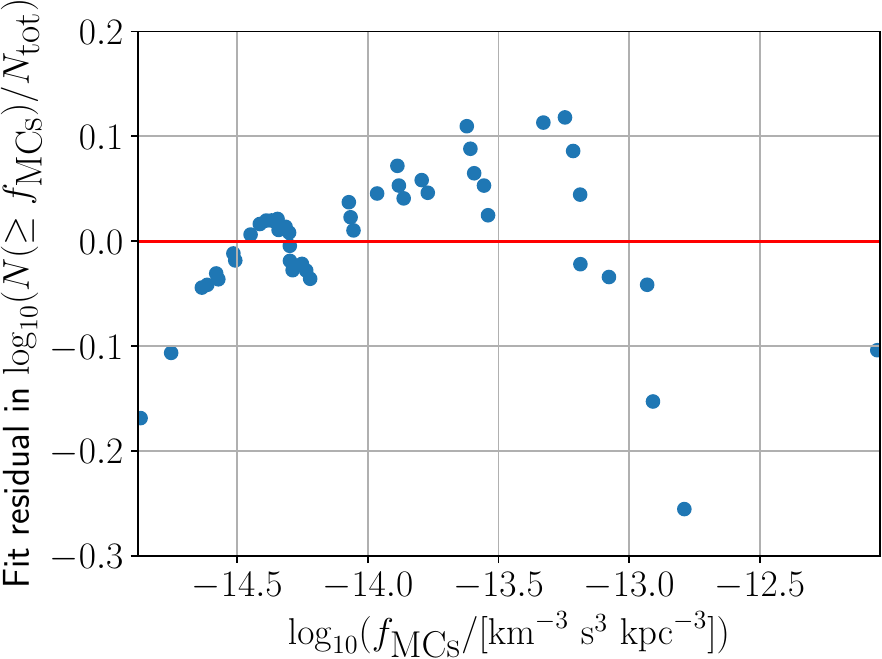}
    \includegraphics[width=8.7cm]{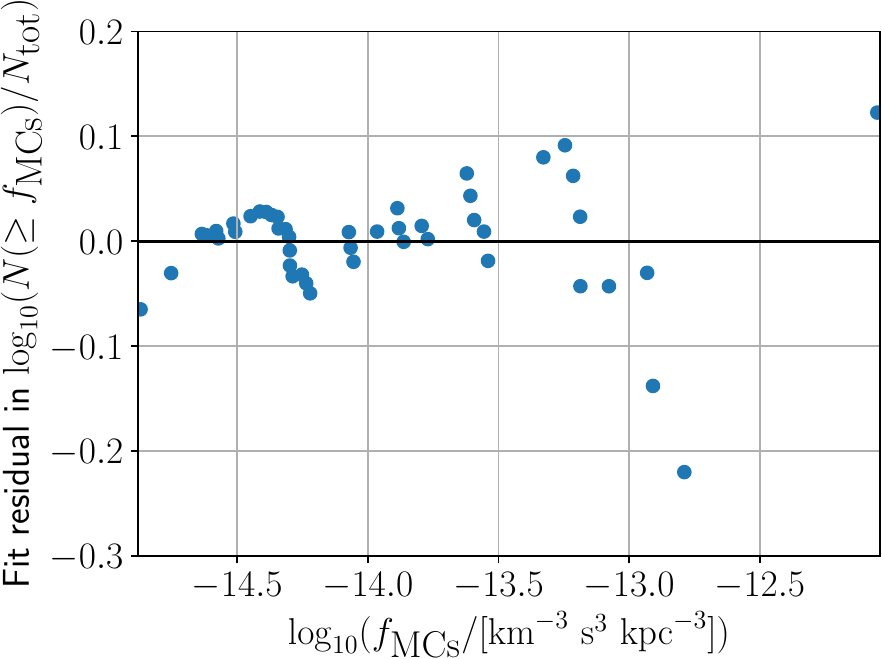}
    \hfill
    \includegraphics[width=8.7cm]{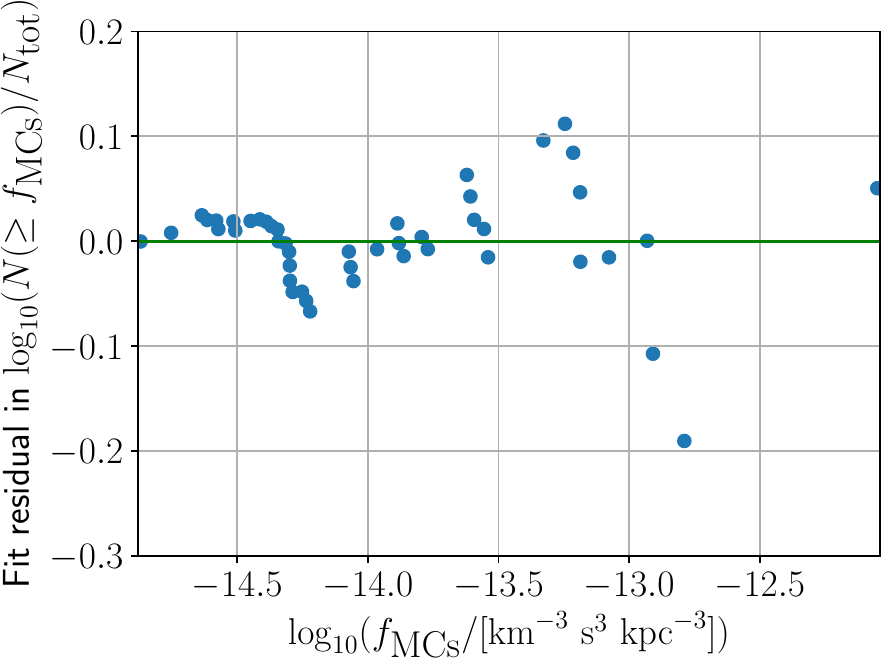}
    \caption{Similar to Figure~\ref{figure_phase_space_extrapolation_TNG50-1}, but now showing the linear (red), quadratic (black), and cubic (green) polynomial fit to the simulated $f_{\mathrm{MCs}}$ distribution in TNG50-1 (\emph{top left}). The other panels show the residuals between the data and the linear (\emph{top right}), quadratic (\emph{bottom left}), and cubic (\emph{bottom right}) fit. In each panel, the gridline at no residual has been emphasized in the same colour as the corresponding fit in the top left panel.}
    \label{figure_polynomial_fits_comparison}
\end{figure*}

In Section~\ref{subsubsec:Phase-space density}, we argued that the distribution of $f_{\mathrm{MCs}}$ in Figure~\ref{figure_phase_space_extrapolation_TNG50-1} is best fit using a quadratic as there is a statistically significant improvement over the linear fit. We also argued that going from a quadratic to a cubic fit does not much improve the fit quality. We illustrate this here in Figure~\ref{figure_polynomial_fits_comparison}, whose top left panel shows all these polynomial fits to the data. The other panels show the residuals between the data and each polynomial fit. It is clear that the residuals are much more tightly clustered around 0 in the quadratic fit compared to the linear fit. However, going further to a cubic fit does not confer any obvious benefit. This justifies our approach of using the quadratic fit in our nominal analysis. We also show results using a linear fit, which should capture the plausible range of uncertainty given that at high $f_{\mathrm{MCs}}$, the cubic fit gives results intermediate between the linear and quadratic fits.

\section{Different selection criteria}
\label{subsec:Different selection criteria}

\begin{figure*}
    \includegraphics[width=5.8cm]{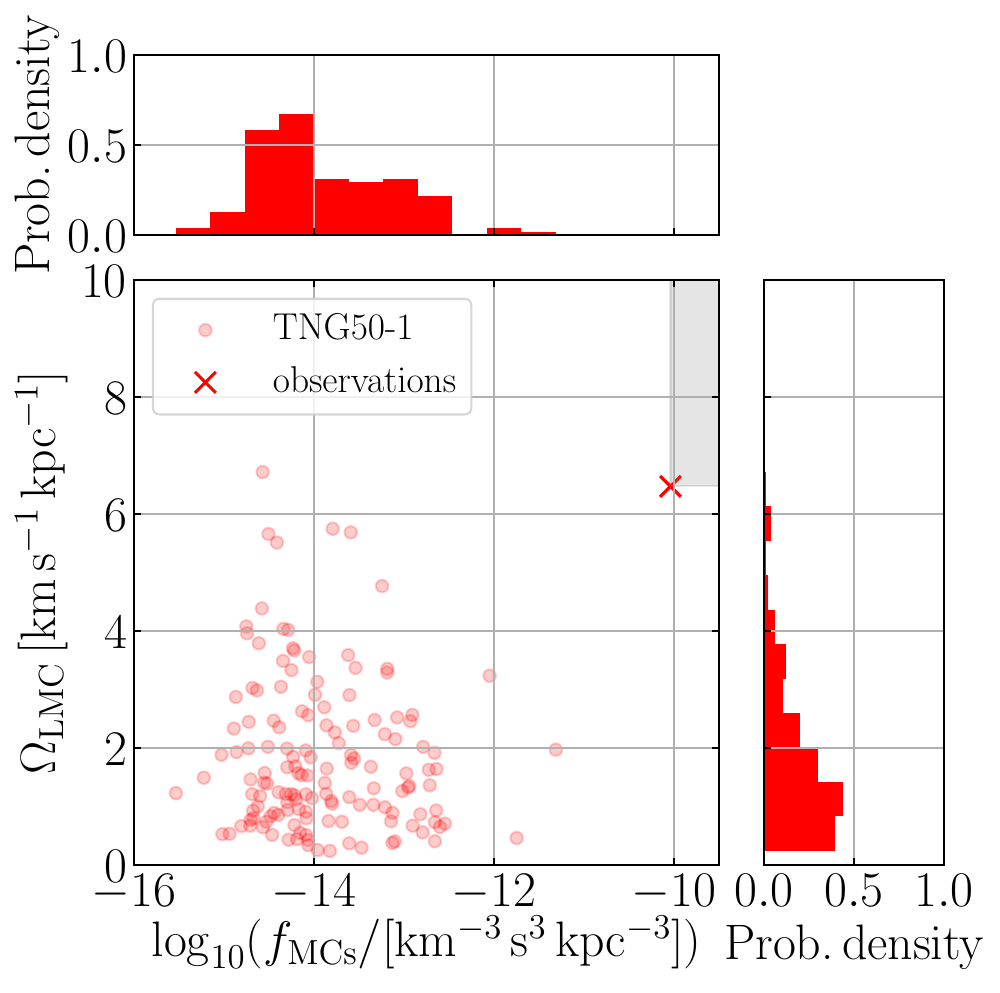}
    \includegraphics[width=5.8cm]{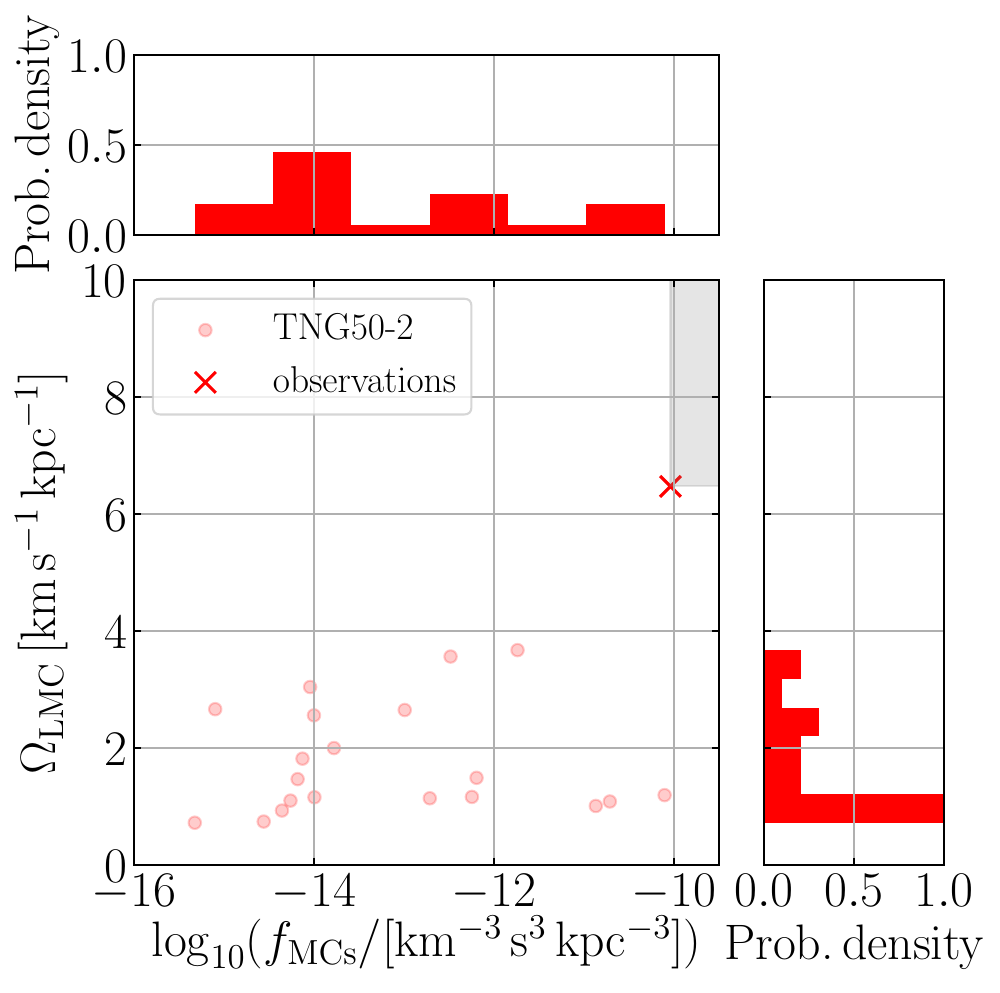}
    \includegraphics[width=5.8cm]{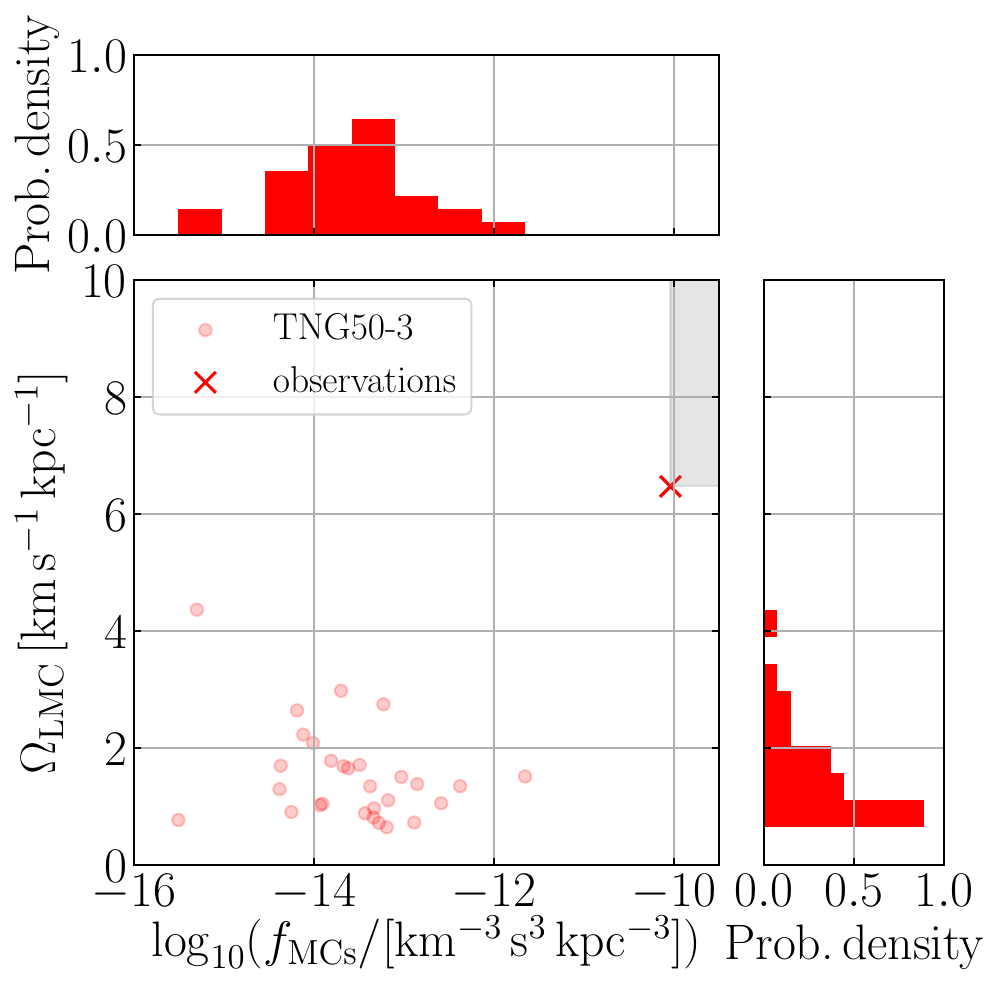}
    \includegraphics[width=5.8cm]{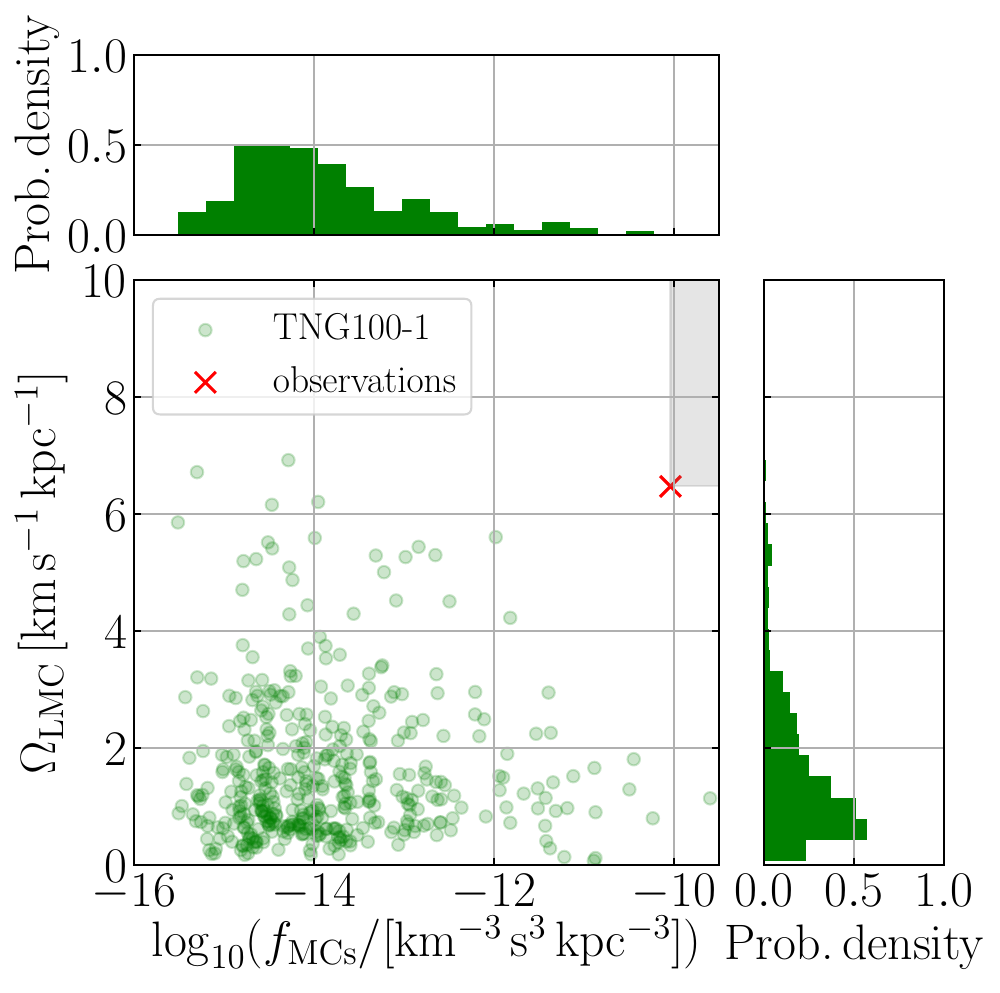}
    \includegraphics[width=5.8cm]{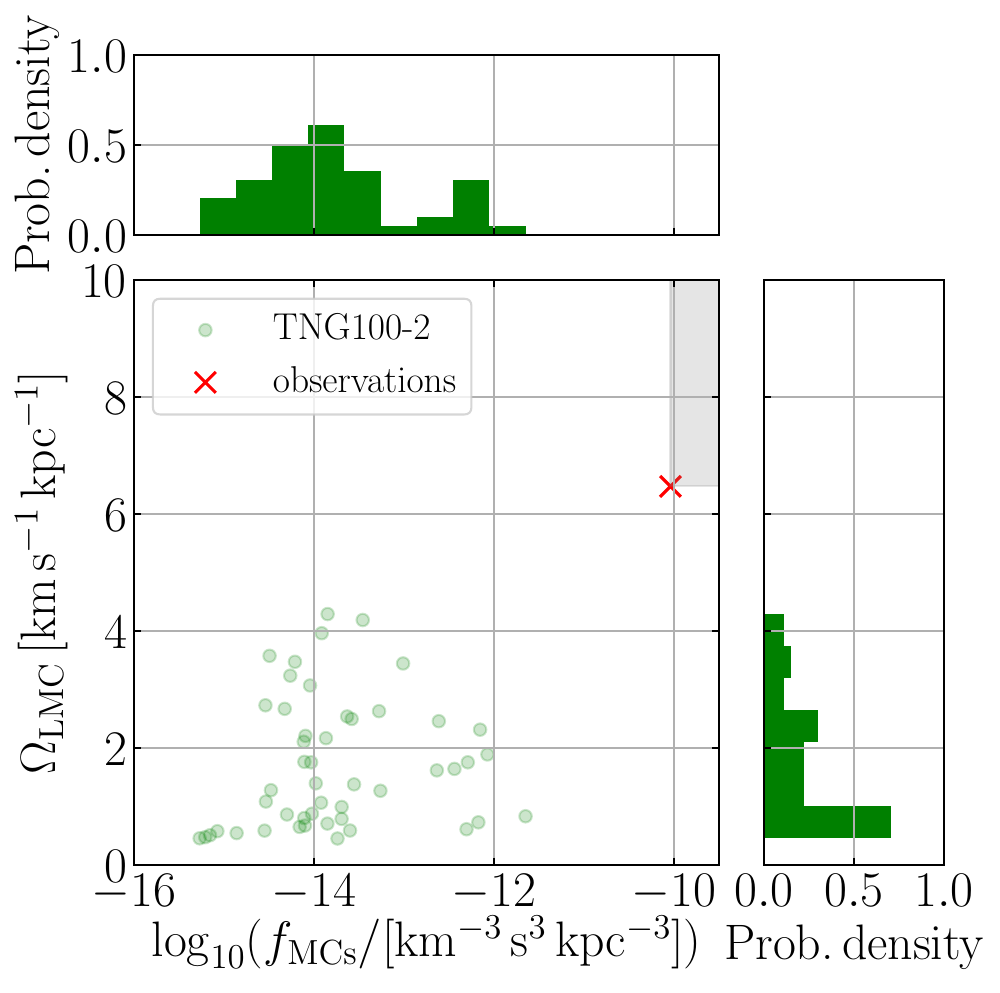}
    \includegraphics[width=5.8cm]{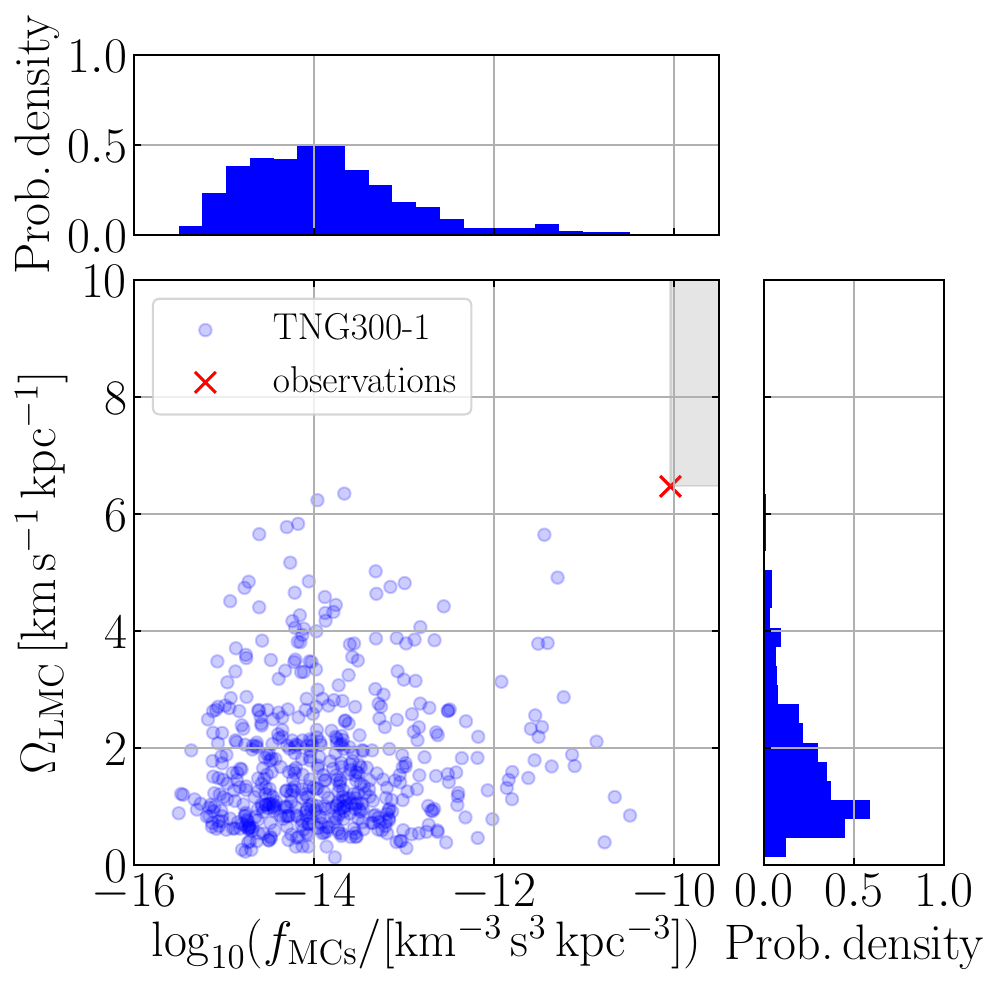}
    \caption{Similar to Figure~\ref{figure_phase_space}, but for all analogues to the MCs with $d_{\mathrm{MW-LMC}} \ge 50 \, \mathrm{kpc}$. None of the $143$ (TNG50-1), $20$ (TNG50-2), $29$ (TNG50-3), $431$ (TNG100-1), $49$ (TNG100-2), or $521$ (TNG300-1) analogues have $f_{\mathrm{MCs}} \ge f_{\mathrm{MCs,obs}}$ and $\Omega_{\mathrm{LMC}} \ge \Omega_{\mathrm{LMC,obs}}$. One of the $1193$ analogues has $f_{\mathrm{MCs}} \ge f_{\mathrm{MCs,obs}}$.}
    \label{figure_phase_space_selected_50kpc}
\end{figure*}

In Section~\ref{subsubsec:Phase-space density} and Appendix~\ref{subsubsec:Effect of resolution}, we focused only on MCs analogues with $50 \, \mathrm{kpc} \le d_{\mathrm{MW-LMC}} \le 100 \, \mathrm{kpc}$. Including also MW-MCs systems with $d_{\mathrm{MW-LMC}} > 100 \, \mathrm{kpc}$ and thus requiring only that $50 \, \mathrm{kpc} \le d_{\mathrm{MW-LMC}} \le 250 \, \mathrm{kpc}$ yields a total sample of 1193 systems in the six TNG runs, as visualized in Figure~\ref{figure_phase_space_selected_50kpc}. Only one of these analogues has $f_{\mathrm{MCs}} \ge f_{\mathrm{MCs,obs}}$ $-$ this system is found in the TNG100-1 run. Taking into account that TNG100-1 has 431 systems with $d_{\mathrm{MW-LMC}} \ge 50 \, \mathrm{kpc}$, this would correspond to a $3.05 \sigma$ tension. The null detection in the TNG50-1 (143 systems) and TNG300-1 (521 systems) runs implies $>2.27\sigma$ and $>3.10\sigma$ tension, respectively. Combining the simulations yields a frequency of 1/1095 ($3.32 \sigma$).

Using the initial sample of MCs analogues in which we require merely that $d_{\mathrm{MW-LMC}} \le 250$~kpc, we find that 3 out of 1312 analogues in the six TNG runs fulfil the phase-space density criterion ($3.05 \sigma$). One of these analogues is found in TNG100-1 (out of 454 systems) and two in TNG300-1 (out of 601 systems), implying a $3.06 \sigma$ and $2.94 \sigma$ tension, respectively. The null detection in TNG50-1 (147 systems) gives a lower limit of $2.71 \sigma$. 

\end{appendix}

\bibliographystyle{aasjournal}
\bibliography{LMC_SMC_LCDM_bbl}
\end{document}